\documentclass[ACS,STIX1COL]{WileyNJD-v2}
\articletype{ORIGINAL ARTICLE}

\usepackage{mathtools}

\mathtoolsset{showonlyrefs=true}

\received{xxx}
\revised{xxx}
\accepted{xxx}

\newcommand{\onlinecite}[1]{\hspace{-1 ex} \nocite{#1}\citenum{#1}} 

\usepackage{bm}

\begin{document}

\title{Non-Markovian quantum kinetic simulations of uniform dense plasmas: mitigating the aliasing problem}

\author[1]{C. Makait}
\author[1]{M. Bonitz}

\authormark{C. Makait and M. Bonitz}

\address[1]{\orgdiv{Institut f\"ur Theoretische Physik und Astrophysik}, \orgname{Christian-Albrechts-Universit\"at zu Kiel}, \orgaddress{\state{Leibnizstra{\ss}e 15, 24098 Kiel}, \country{Germany}}}

\corres{*\email{makait@physik.uni-kiel.de}}

\abstract{
Dense quantum plasmas out of equilibrium are successfully modeled using quantum kinetic equations, such as the quantum Boltzmann, Landau or Balescu-Lenard equation. However, these equations do not properly take into account correlation effects which requires to use generalized non-Markovian kinetic equations. While the latter have been successful for lattice models, applications to continuous systems such as plasmas are severely hampered by aliasing effects. Here we present a strategy how to suppress aliasing and make applications of non-Makrovian quantum kinetic equations to plasmas possible.
}


\maketitle

\section{Introduction}\label{sec:1}
Recently there has been rapidly growing interest in warm dense matter (WDM), which is driven in particular by the progress in inertial confinement fusion research \cite{icf-collab_prl_22, icf-collab_prl_24,pak-icf_pre_24}. Theoretical modeling of WDM is complicated due to the need to simultaneously take into account Coulomb interaction and electronic quantum and spin effects. There is a broad arsenal of computational techniques available to compute the plasma properties in thermodynamic equilibrium,  which include quantum Monte Carlo (QMC) simulations, density functional theory (DFT), and hydrodynamic models, for recent overviews see \cite{dornheim_physrep_18,bonitz_pop_20, bonitz_pop_24}. In recent years, plasmas driven out of equilibrium e.g., by the application of high-intensity lasers, attracted high interest and many theoretical works appeared that studied the temperature dynamics, the dynamic structure factor or optical properties, e.g.
\cite{gericke_prl11, medvedev_prl_11, kluge_prl_11, vorberger_pre18}.
In particular, the electron distribution functions may strongly differ from a Fermi function. In that case it is important to compute the nonequilibrium momentum distributions and its time evolution which is governed by quantum kinetic equations \cite{bonitz_98teubner,bonitz_aip_12}. There is abundant literature on the solution of quantum kinetic equations for dense plasmas. For example, Kosse et al. solved the quantum Landau equation \cite{kosse-etal.97} that uses a statically screened Coulomb potential and applies to weak coupling. To properly take into account dynamical screening effects one has to resort to the quantum Balescu-Lenard equation \cite{balescu60,lenard60} that was first solved for electron-hole plasmas by Scott et al.~\cite{scott_prl_92,binder_prb_92}, and for plasmas by Bonitz et al.~\cite{bonitz-etal.96ib2} and Scullard et al. \cite{graziani_pop_13}. These equations involve a kinetic energy conserving delta function (Fermi's golden rule) and, thus, violate the correct conservation law of total (kinetic plus interaction) energy. Moreover, they fail to correctly describe the short-time behavior, for a discussion, see e.g., Ref.~\onlinecite{bonitz_pop_24}. 

These shortcomings were overcome by resorting to generalized quantum kinetic equations that have been derived in the framework of reduced density operators \cite{bonitz-etal.96jpcm,bonitz-etal.96pla,kremp-etal.97ap,bonitz_98teubner} or nonequilibrium Green functions (NEGF), e.g.,\cite{green-book,haug_2008_quantum,schluenzen_jpcm_19}, and first solutions for uniform electron-hole plasmas were presented in Refs.~\onlinecite{bonitz-etal.96jpcm,binder-etal.97prb,kwong-etal.98pss} followed by applications to dense quantum plasmas \cite{bonitz-etal.97pre,semkat_99_pre,semkat_00_jmp,kwong_prl_00} and dense laser plasmas \cite{kremp_99_pre,bonitz_99_cpp,haberland_01_pre}
by Kremp, Bonitz and co-workers.
Note that the generalized quantum kinetic equations involve a memory integral \cite{bonitz_98teubner}, making their numerical solution computationally costly: the CPU time scales quadratically with the number of time steps, $N_t$, or even cubically, if the equations of motion for the two-time NEGF are being solved. This problem could recently be overcome with the G1-G2 scheme that has linear scaling with $N_t$ \cite{schluenzen_prl_20} which was demonstrated for weak and strong coupling (generalized quantum Landau and Boltzmann equation, respectively), for dynamical screening (generalized Balescu-Lenard equation corresponding to the GW approximation) \cite{joost_prb_20} and even for the simultaneous account of strong coupling and dynamical screening (dynamically screened ladder approximation) \cite{joost_prb_22}, as well as for the coupling to bosons \cite{karlsson_prl21}. So far, the G1-G2 scheme was applied to finite discrete basis sets. such as lattice models or small molecules \cite{schluenzen_prl_20, joost_prb_20, pavlyukh_prb_22, pavlyukh_time-linear_2022}. 

However, for the description of dense nonequilibrium plasmas, the proper choice is a basis of momentum states (plane waves) that sufficiently well cover the argument of the momentum distribution function $f(\textbf{p},t)$. This was the obvious choice in previous quantum kinetic simulations with Markovian kinetic equations, e.g., \cite{kosse-etal.97, kwong_prl_00, graziani_pop_13} which afforded a sufficiently large momentum grid. However, within the G1-G2 scheme, coupled equations for the single-particle and two-particle Green functions, $G$ and $\mathcal{G}$ are being solved where the latter depends on three momenta, $\mathcal{G} = \mathcal{G}(\textbf{p}_1,\textbf{p}_2,\textbf{p}_3,t)$. This imposes severe restrictions on the number of grid points that can be afforded. Since often $\mathcal{G}$ is strongly varying in momentum space, this gives rise to so-called ``aliasing'' effects resulting in unstable quantum kinetic simulations, for long times.
The goal of this paper is to analyze the aliasing problem in detail and to develop computational strategies that eliminate aliasing effects and make quantum kinetic simulations for dense plasmas feasible.

This paper is structured as follows. In Sec.~\ref{sec:2} we introduce the method of nonequilibrium Green functions and formulate the equations of motion for the two-time NEGF in momentum representation. The transition to time-diagonal quantum kinetic equations is performed in Sec.~\ref{sec:gkba}. There we demonstrate the aliasing problem and introduce a solution via a diffusion approach. Numerical results are presented in Sec.~\ref{sec:3} and a conclusion and outlook will be given in Sec.~\ref{sec:4}.

\section{Theory}\label{sec:2}
\subsection{Uniform system Hamiltonian}
For the quantum kinetic theory of spatially uniform plasmas the natural choice is the momentum representation where the basis states are the momentum-spin eigenstates, $|i\rangle \longrightarrow |\mathbf{p}\sigma\rangle$ with $\langle \mathbf{r}\alpha|\mathbf{p}\sigma\rangle=\delta_{\alpha\sigma}e^{i\mathbf{p}\cdot\mathbf{r}}\frac{1}{\sqrt{L}^d}$. Here, $d$ denotes the dimension of the system and $L$ the length of the box. In a multi-component system, the species index mostly behaves like a spin index in regard to the structure of the equations, hence greek spin indices can also be interpreted as to also describe different particle species, e.g., ions. 
In a uniform system consisting of electrons in front of a neutralizing background, we can write
\begin{align}
  \hat{H}=\hat{H}_{\text{ee}}+\hat{H}_{be}+H_{bb},
\end{align}
where the electronic part $\hat{H}_{\text{ee}}$ is given by
\begin{align}
    \hat{H}_{\text{ee}}=\sum\limits_{\textbf{k},\sigma}\frac{\hbar^2_{}\textbf{k}^2}{2m_{\text{e}}}
    \hat{a}_{\textbf{k}\sigma}^{\dagger}
    \hat{a}_{\textbf{k}^{}\sigma}+
    \frac{1}{2}\sum\limits_{\textbf{k}\textbf{k}'\textbf{q}\atop \sigma\sigma'}w(q,t)
    \hat{a}_{\textbf{k}+\textbf{q},\sigma}^{\dagger}
    \hat{a}_{\textbf{k}'-\textbf{q},\sigma'}^{\dagger}
    \hat{a}_{\textbf{k}'\sigma'}^{}
    \hat{a}_{\textbf{k}\sigma}^{}\,,
\label{HEG_hamiltonian2}
\end{align}
where $\hat{a}_{\textbf{k}+\textbf{q},\sigma}^{\dagger}$ and $\hat{a}_{\textbf{k}^{}\sigma}$ are standard second quantization creation and annihilation operators that obey commutation (anticommutation) relations for bosons (fermions).
The remaining terms, $\hat{H}_{\text{be}}$ and $H_{\text{bb}}$, describe the electron-background and the background-background interaction. It turns out ('Coulomb-regularization') that they exactly compensate the $\mathbf{q}=0$ part in $\hat{H}_{\text{ee}}$, and therefore we can simply write
\begin{align}
    \hat{H}=\sum\limits_{\textbf{k},\sigma}\frac{\hbar^2_{}\textbf{k}^2}{2m_{\text{e}}}
    \hat{a}_{\textbf{k}\sigma}^{\dagger}
    \hat{a}_{\textbf{k}^{}\sigma}+
    \frac{1}{2}\sum\limits_{\textbf{k}\textbf{k}',\textbf{q}\neq 0\atop \sigma\sigma'}w(q,t)
    \hat{a}_{\textbf{k}+\textbf{q},\sigma}^{\dagger}
    \hat{a}_{\textbf{k}'-\textbf{q},\sigma'}^{\dagger}
    \hat{a}_{\textbf{k}'\sigma'}^{}
    \hat{a}_{\textbf{k}\sigma}^{}\,.
\end{align}

The matrix elements of Coulomb and Yukawa potentials can be computed in 2D and 3D, whereas in 1D, they diverge. The matrix elements in 3D are given by
\begin{align}
   \left\langle\textbf{k}-\mathbf{q},\alpha;\textbf{p}+\mathbf{q},\beta\left| Z_\alpha Z_\beta e^2\frac{e^{-\kappa r}}{r} \right|\textbf{k}\alpha;\textbf{p}\beta\right\rangle=  Z_\alpha Z_\beta w\left(q\right)=\frac{4\pi Z_\alpha Z_\beta e^2 }{q^2+\kappa^2}\,.
\label{eq:3D-pair-potential-fourier}
\end{align}
where $Z_\alpha,Z_\beta$ are the charge numbers. The Coulomb and Yukawa matrix elements can be computed in so-called quasi-1D systems, which means that it is embedded in 3D but is confined in two dimensions. In a harmonic confinement, the states assume the form\cite{giuliani_quantum_2005}
\begin{align}
   \langle\mathbf{r}\alpha'|k\alpha\rangle = \frac{1}{\sqrt{L}}\left(\frac{2}{\pi a^2}\right)^\frac{1}{2}\exp{\left(-\frac{r_\perp^2}{a^2}\right)}\exp{\left(\mathrm{i}r_\parallel\cdot k\right)}\delta_{\alpha\alpha'},
   \label{eq:1p-states}
\end{align}
where the length $a$ defines the radius of the ground state of the confinement potential. Once the form of the state is specified, we can also compute Coulomb or Yukawa matrix elements in quasi-1D,
\begin{align}
   \left\langle k-q,\alpha;p+q,\beta\left| Z_\alpha Z_\beta e^2\frac{e^{-\kappa r}}{r} \right|k\alpha;p\beta\right\rangle=   Z_\alpha Z_\beta w\left(q\right)=-Z_\alpha Z_\beta e^{2}\exp\left[(q^{2}+\kappa^{2})a^{2}\right]\text{Ei}\left(-(q^{2}+\kappa^{2})a^{2}\right),
\label{eq:pair-potential-fourier}
\end{align}
which will be used in our example calculations. 

\subsection{Nonequilibrium Green functions}\label{sec:TheoryNEGF}
The Nonequilibrium Green functions (NEGF) is defined as the expectation value of creation and annihilation operators in the Heisenberg picture for which there exist two independent versions (correlation functions),
\begin{align}
    G^<_{ij}(t,t')=\pm\frac{1}{\mathrm{i}\hbar}\left\langle \hat{a}_{j}^\dagger(t')\,\hat{a}_i(t)\right\rangle,\qquad\qquad G^>_{ij}(t,t')=\frac{1}{\mathrm{i}\hbar}\left\langle \hat{a}_i(t)\,\hat{a}_{j}^\dagger(t')\right\rangle,
    \label{eq:ggtrless-def}
\end{align}
where the upper (lower) sign corresponds to bosons (fermions). In a uniform system, in the momentum basis, the Green functions are diagonal in the momentum-spin index,
\begin{align}
 G^\gtrless_{\mathbf{p}\alpha,\mathbf{p}'\alpha'}(t,t')=\vcentcolon G^\gtrless_{\mathbf{p}\alpha}(t,t')\delta_{\mathbf{p},\mathbf{p}'}\delta_{\alpha,\alpha'}\,,
\end{align}
The NEGF determines all time-dependent expectation values of single-particle observables via
\begin{align}
    \langle \hat{O}\rangle(t)=\sum\limits_{\mathbf{p}\sigma}f_{\mathbf{p}\sigma}(t)\langle \mathbf{p}\sigma |\hat{O}^{(1)}|\mathbf{p}\sigma\rangle \equiv \sum\limits_{\mathbf{p}\sigma}f_{\mathbf{p}\sigma}(t)O_{\mathbf{p}\sigma}\,,
\end{align}
where the single-particle distribution function follows from the NEGF via $f_{\mathbf{p}\sigma}(t)=\pm i\hbar G^<_{\mathbf{p}\sigma}(t,t)$. 
The time-dependence along the time diagonal of $G^\gtrless(t,t)$ follows from the time-diagonal Keldysh-Kadanoff-Baym equations (KBE),
\begin{align}
    i\hbar\frac{d}{dt}G^\gtrless_{\mathbf{p}\sigma}(t,t)&=I_{\mathbf{p}\sigma}(t)=\int_{t_0}^t \left\{\Sigma^>_{\mathbf{p}\sigma}(t,\bar t)G^<_{\mathbf{p}\sigma}(\bar t,t)-\Sigma^<_{\mathbf{p}\sigma}(t,\bar t)G^>_{\mathbf{p}\sigma}(\bar t,t)\right\}d\bar t + c.c.\,,\label{eq:TimeDiagKBE}
\end{align}
where $\Sigma^\gtrless_{\mathbf{p}\sigma}(t,t')$ is the many-body selfenergy which contains all correlation effects. Its exact form is not known, so approximations have to be used; the most common ones will be discussed in Sec. \ref{sec:TheorySelfenergy}. Since the selfenergy is the key quantity that gives access to correlations, the total interaction energy has a contribution depending on $\Sigma^\gtrless$. The interaction energy consists of exchange energy and correlation energy, since the Hartree energy is cancelled exactly by the neutralizing background. This leaves us with the following expressions for the energy:
\begin{align}
    E_\mathrm{int}(t)&=E_\mathrm{Fock}(t) + E_\mathrm{corr}(t)\\
    E_\mathrm{Fock}(t)&=\frac{-i\hbar}{2} \sum\limits_{\mathbf{p}\sigma} U^\mathrm{F}_{\mathbf{p}\sigma}(t)G^<_{\mathbf{p}\sigma}(t,t)\,,\qquad\text{where}\qquad U^\mathrm{F}_{\mathbf{p}\sigma}(t)=i\hbar Z_\sigma^2\sum\limits_{\mathbf{q}}w_\mathbf{q}(t)G^<_{\mathbf{p}-\mathbf{q},\sigma}(t,t)\\
    E_\mathrm{corr}(t)&=-\frac{1}{2}\mathfrak{Im}\Big\{\sum\limits_{\mathbf{p}\sigma} I_{\mathbf{p}\sigma}(t)\Big\}
\end{align}

\subsection{Selfenergy approximations}\label{sec:TheorySelfenergy}
In uniform systems out of equilibrium, the most commonly used selfenergy approximations are the direct Second-Order Born approximation (SOA) and the $GW$ approximation. In both cases, the selfenergy is given by a convolution in momentum space. For GW, we have,
\begin{align}
    \Sigma^\gtrless_{\mathbf{p}\sigma}(t,t')=i\hbar Z_{\sigma}^2 \sum\limits_{\mathbf{p}}W^\gtrless_{\mathbf{q}}(t,t')G^\gtrless_{\mathbf{p}-\mathbf{q},\sigma}(t,t')\,,\label{eq:SigmaGW}
\end{align}
where the screened interaction, $W$, is the solution of a Dyson equation,
\begin{align}
    W^\gtrless_\mathbf{q}(t_1,t_2) =&\pi^\gtrless_\mathbf{q}(t_1,t_2)\,w_\mathbf{q}(t_1)\,w_\mathbf{q}(t_2)+w_\mathbf{q}(t_1)\int\limits_{t_0}^{t_1}\left[\pi^{>}_\mathbf{q}(t_1,\Bar{t})-\pi^{<}_\mathbf{q}(t_1,\Bar{t})\right]\,W^\gtrless_\mathbf{q}(\Bar{t},t_2)\ \text{d}\Bar{t} + \\
    &+ w_\mathbf{q}(t_1)\int\limits_{t_0}^{t_2}\pi^\gtrless_\mathbf{q}(t_1,\Bar{t})\,\left[W^{<}_\mathbf{q}(\Bar{t},t_2)-W^{>}_\mathbf{q}(\Bar{t},t_2)\right]\ \text{d}\Bar{t}\,,\label{eq:DysonWGLUniform2}
\end{align}
with the polarization functions being
 given by another convolution [the sign $+$ ($-$) refers to bosons (fermions).],
\begin{align}
    \pi^\gtrless_\mathbf{q}(t_1,t_2)&=\mathrm{i}\hbar\sum\limits_{\mathbf{k}'\beta} (\pm)_\beta Z_\beta^2 G^\gtrless_{\mathbf{k}'+\mathbf{q},\beta}(t_1,t_2)\,G^\lessgtr_{\mathbf{k}',\beta}(t_2,t_1)\,.\label{eq:defPolarizationPi2}
\end{align}
If the integral terms on the r.h.s. of Eq.~(\ref{eq:DysonWGLUniform2}) are omitted, one recovers the direct SOA selfenergy. 

In a spatially uniform system, the mean-field does not induce any dynamics. Hence, relaxation only appears once correlations are included. The direct SOA selfenergy is the simplest approximation for correlations. There exists a second SOA term that is of second order in the interaction as well -- the exchange diagram -- that is not of convolution structure -- and is therefore, much more difficult to compute. The $GW$ approximation retains the convolution structure, but contains screening dynamics via the Dyson equation of $W$. Among its included physical effects are dynamical screening, plasmons, plasma instabilities, and screening build-up. Since these effects play an important role in many plasmas, the $GW$ approximation is a popular choice, albeit computationally more expensive than the direct SOA.

Another important selfenergy choice are the (statically screened) ladder approximations (T-Matrix in particle-particle and particle-hole channel), which excel at the description of strong coupling. Strong coupling is especially important in the low-density limit where the dynamics are predominantly determined by binary collisions. 
Since the solution of the time-dependent Lippmann--Schwinger equation for the T-matrix is difficult, nonequilibrium solutions have been primarily reported for lattice systems, whereas for the uniform gas, most results were restricted to equilibrium, to the Markov limit (Boltzmann equation). Finally, the $GW$ and the ladder approximations can be combined in an approximate way within the Gould-DeWitt scheme \cite{GouldDeWitt67}. A selfconsistent non-Markovian combination is possible with the G1--G2 scheme and yields the nonequilibrium \textit{Dynamically Screened Ladder} Approximation \cite{joost_prb_22}.

\section{Single-time quantum kinetic equations and the aliasing problem}\label{sec:gkba}
\subsection{Generalized Kadanoff--Baym ansatz and G1--G2 scheme}\label{sec:TheoryG1G2}
The collision integral for the time-diagonal (single-time) NEGF, Eq. \eqref{eq:TimeDiagKBE}, is dependent on the two-time dependent NEGF, both explicitly, as seen in the formula \eqref{eq:TimeDiagKBE}, and implicitly, as $\Sigma_{\mathbf{p}\sigma}(t,t')$ is dependent on the whole history of $G$. One way to close the equation of motion for $G(t,t)$ is to use the Generalized Kadanoff--Baym ansatz (GKBA)\cite{lipavski_prb_86}, 
\begin{align}
    G^\gtrless_{\mathbf{p}\sigma}(t_1,t_2)=-\mathrm{i}\hbar  \left(G^{\mathcal{R}}_{\mathbf{p}\sigma}(t_1,t_2)\,G^\gtrless_{\mathbf{p}\sigma}(t_2)-G^\gtrless_{\mathbf{p}\sigma}(t_1)\,G^{\mathcal{A}}_{\mathbf{p}\sigma}(t_1,t_2)\right)\,,
\end{align}
which contains the time-diagonal functions $G^\gtrless_{\mathbf{p}\sigma}(t) \equiv G^\gtrless_{\mathbf{p}\sigma}(t,t)$, but still involves two-time retarded and advanced Green functions. To avoid the solution of two-time equations, the propagators are approximated by
\begin{align}
    G^{\mathcal{R}/\mathcal{A},\text{HF}}_{\mathbf{p}\sigma}(t_1,t_2)&=\pm\frac{1}{\mathrm{i}\hbar}\Theta\left(\pm [t_1-t_2]\right)\,\exp\left\{\frac{1}{\mathrm{i}\hbar}\int_{t_2}^{t_1}\text{d}\bar{t}\, 
    h^{\rm HF}_{\textbf{p},\sigma}(\bar t)
    \,\right\}\,,\label{eq:gra-hf-def}\\
    h^{\rm HF}_{\textbf{p},\alpha}(t) &= E^\alpha(\textbf{p}) + U^F_{\mathbf{p}\sigma}({t})\,, \quad E^\alpha(\textbf{p}) = \frac{\textbf{p}^2}{2m_\alpha}\,,
    \label{eq:h-hf-def}
\end{align}
which is the \textit{Hartree--Fock GKBA}. If, further, the mean field terms are neglected, $U^F\to 0$, this reduces to the \textit{free GKBA}, which yields the dominant contribution in the weak-coupling limit. For a detailed discussion of the HF-GKBA and its properties we refer to Ref.~\onlinecite{bonitz_pssb23}.
The main advantage of the HF-GKBA is that it allows for a time-local reformulation -- the \textit{G1--G2 scheme}\cite{schluenzen_prl_20} -- which reduces the CPU scaling from $\mathcal{O}(N_t^3)$ to $\mathcal{O}(N_t)$ and the memory demand from $\mathcal{O}(N_t^2)$ to $\mathcal{O}(1)$. This makes the G1--G2 scheme viable for long simulation times. Within the G1--G2 scheme, even the non-Markovian T-Matrix approximations come within reach for systems with long-range interaction. The main bottleneck of the scheme is the need to store and propagate a $2$-particle quantity, $\mathcal{G}(t)$
 -- the correlated part of the two-time Green function.

 The two-particle Green function is particularly costly in the momentum representation:
$\mathcal{G}^{\alpha\beta}_{\mathbf{kpq}}(t)$ is depending on three momentum vectors, in a uniform system. The three momenta indexing two-particle quantities are a shortened form for
\begin{align}
    X_{\mathbf{kpq}}^{\alpha\beta}\equiv X^{\alpha\beta\alpha\beta}_{\mathbf{k}+\mathbf{q},\mathbf{p}-\mathbf{q},\mathbf{k},\mathbf{p}}
\end{align}
$\mathcal{G}^{\alpha\beta}_{\mathbf{kpq}}(t)$ is defined by the property \cite{schluenzen_prl_20}
\begin{align}
    I_{\mathbf{p}\sigma}(t)=\int_{t_0}^t \left\{\Sigma^>_{\mathbf{p}\sigma}(t,\bar t)G^<_{\mathbf{p}\sigma}(\bar t,t)-\Sigma^<_{\mathbf{p}\sigma}(t,\bar t)G^>_{\mathbf{p}\sigma}(\bar t,t) \right\} d\bar t =-i\hbar \sum\limits_{\mathbf{kq},\alpha}Z_\alpha Z_\sigma w_\mathbf{q}(t)\mathcal{G}^{\alpha \sigma}_{\mathbf{kpq}}(t)\,,\label{eq:col-int-g2}
\end{align}
and obeys an ordinary differential equation\cite{bonitz_springer_10,schluenzen_prl_20,joost_prb_20}, if one uses e.g., second-order, $GW$, or T-Matrix selfenergies. 
In the case of $GW$ it is given by 
\begin{align}
    \mathrm{i}\hbar \frac{\text{d}}{\text{d}t}\mathcal{G}^{\alpha\beta}_{\mathbf{kpq}}(t)-\left[h^{\text{HF},(2)}(t),\mathcal{G}(t)\right]_{\mathbf{kpq}}^{\alpha\beta}&=\Psi^{\alpha\beta}_{\mathbf{kpq}}(t)+\Pi^{\alpha\beta}_{\mathbf{kpq}}(t)\,,\label{eq:G2GW}
\end{align}
with the definitions
\begin{align}
    \left[h^{\text{HF},(2)}(t),\mathcal{G}(t)\right]_\mathbf{kpq}^{\alpha\beta}&=\mathcal{G}^{\alpha\beta}_{\mathbf{kpq}}(t)\left(h^\text{HF}_{\mathbf{k}+\mathbf{q},\alpha}(t)+h^\text{HF}_{\mathbf{p}-\mathbf{q},\beta}(t)-h^\text{HF}_{\mathbf{k},\alpha}(t)-h^\text{HF}_{\mathbf{p},\beta}(t)\right)\,,
    \label{eq:2pCommutatorMomRep}
\end{align}
and
\begin{align}
    \Psi^{\alpha\beta}_{\mathbf{kpq}}(t)&=(\mathrm{i}\hbar)^2 w_{|\mathbf{q}|}^{\alpha\beta}(t)\left(G^>_{\mathbf{k}+\mathbf{q},\alpha}(t)\,G^>_{\mathbf{p}-\mathbf{q},\beta}(t)\,G^<_{\mathbf{k},\alpha}(t)\,G^<_{\mathbf{p},\beta}(t)-G^<_{\mathbf{k}+\mathbf{q},\alpha}(t)\,G^<_{\mathbf{p}-\mathbf{q},\beta}(t)\,G^>_{\mathbf{k},\alpha}(t)\,G^>_{\mathbf{p},\beta}(t)\right)\,,\label{eq:defPsi}\\
    \Pi^{\alpha\beta}_{\mathbf{kpq}}(t)&=\pi_{\mathbf{kpq}}^{\alpha\beta}(t)-\left[\pi_{\mathbf{p}-\mathbf{q},\mathbf{k}+\mathbf{q},\mathbf{q}}^{\beta\alpha}(t)\right]^*,\quad\text{where}\quad \pi_{\mathbf{kpq}}^{\alpha\beta}= (\mathrm{i}\hbar)^2\left[G^>_{\mathbf{p}-\mathbf{q},\beta}(t)\,G^<_{\mathbf{p},\beta}(t)-G^<_{\mathbf{p}-\mathbf{q},\beta}(t)\,G^>_{\mathbf{p},\beta}(t)\right]\sum\limits_{\mathbf{p}'\gamma}w_{|\mathbf{q}|}^{\alpha\gamma}(t)\,\mathcal{G}^{\alpha\gamma}_{\mathbf{k}\mathbf{p}'\mathbf{q}}(t)\,,
    \nonumber
\end{align}
with the short notation $w^{\alpha\beta}_\mathbf{q}=Z_\alpha Z_\beta w_\mathbf{q}.$ The function $\mathcal{G}$ is related to the time-diagonal two-particle Green function and the two-particle reduced density matrix $F^{(2)}$ by
\begin{align}
    F^{(2),\alpha\beta}_{\mathbf{k}\mathbf{p}\mathbf{q}}(t)&=(i\hbar)^2(\pm)_\alpha (\pm)_\beta\left[G^<_{\mathbf{k}\alpha}(t)\,G^<_{\mathbf{p}\beta}(t)\,\delta_{\mathbf{q}0} \pm G^<_{\mathbf{k}\alpha}(t)\,G^<_{\mathbf{p}\beta}(t)\,\delta_{\mathbf{k}+\mathbf{q}-\mathbf{p},0}\delta_{\alpha\beta}+\mathcal{G}^{\alpha\beta}_{\mathbf{kpq}}(t)\right]\,.
\end{align}
Thus, in addition to simplifying the propagation of $G^\gtrless_{\mathbf{p}\sigma}(t,t)$, the G1--G2 scheme also gives access to two-particle observables, such as  the pair distribution and the static structure factor.
The equation of motion for $\mathcal{G}$ in SOA (with direct and exchange terms) is obtained from Eq. \eqref{eq:G2GW} by neglecting the $\Pi$ term and replacing $w^{\alpha\beta}_{\mathbf
{q}}(t)$ in Eq. \eqref{eq:defPsi} by the (anti-)symmetrized interaction, $w^{\alpha\beta}_{\mathbf
{q}}(t)\pm \delta_{\alpha\beta}w_{\mathbf{k}+\mathbf{q}-\mathbf{p}}^{\alpha\beta}(t)$. Without the exchange term ($\pm$), this corresponds to the direct SOA.
Equation \eqref{eq:G2GW}, together with the single-particle equation
\begin{align}
    i\hbar\frac{d}{dt}G^\gtrless_{\mathbf{p}\sigma}(t)&=I_{\mathbf{p}\sigma}(t)\,,\label{eq:g1-equation}
\end{align}
constitutes the momentum representation of the G1-G2 scheme for spatially uniform systems~\cite{schluenzen_prl_20}.

The second, equivalent form of the HF-GKBA with SOA selfenergies is obtained by a formal solution of Eq.~\eqref{eq:G2GW} which yields a closed expression for the single-particle NEGF, Eq.~\eqref{eq:g1-equation}, with the explicit result for the collision integral 
\begin{align}
    I_{\mathbf{p}\alpha}(t) = &(i\hbar)^2Z_\alpha^2\sum\limits_{\beta} (\pm)_\beta Z_\beta^2\int_{0}^{t-t_0}\text{d}\tau\int\frac{\text{d}\mathbf{p}_2}{(2\pi\hbar)^d}\int\frac{\text{d}\mathbf{q}}{(2\pi\hbar)^d}w_{\mathbf{q}}\left[w_\mathbf{q}\pm \delta_{\alpha\beta}w_{\mathbf{p}-\mathbf{p}_2-\mathbf{q}}\right]\exp\left[\frac{-i}{\hbar}\int_{t-\tau}^t \left( h^{\text{HF}}_{\mathbf{p},\alpha}+h^{\text{HF}}_{\mathbf{p}_2,\beta}-h^{\text{HF}}_{\mathbf{p}+\mathbf{q},\alpha}-h^{\text{HF}}_{\mathbf{p}_2-\mathbf{q},\beta} \right) d\bar{t}\right)\nonumber\label{eq:CollIntSOAMemory}\\
    &\times \left[G^<_{\mathbf{p}+\mathbf{q},\alpha}(t-\tau)\,G^<_{\mathbf{p}_2-\mathbf{q},\beta}(t-\tau)\,G^>_{\mathbf{p},\alpha}(t-\tau)\,G^>_{\mathbf{p}_2,\beta}(t-\tau)-G^>_{\mathbf{p}+\mathbf{q},\alpha}(t-\tau)\,G^>_{\mathbf{p}_2-\mathbf{q},\beta}(t-\tau)\,G^<_{\mathbf{p},\alpha}(t-\tau)\,G^<_{\mathbf{p}_2,\beta}(t-\tau)\right]\,.
\end{align}
where the Hartree--Fock Hamiltonians $h^\mathrm{HF}$ are evaluated at time $\bar{t}.$ This is the standard non-Markovian version of a quantum kinetic equation following from the HF-GKBA with SOA selfenergies. In the following we will discuss aliasing effects in both, the time local and the non-Markovian versions of the HF-GKBA.

\subsection{Aliasing effects in the G1-G2 scheme}\label{sec:TheoryAliasing}
\begin{figure}
    \centering
    \includegraphics[width=\linewidth]{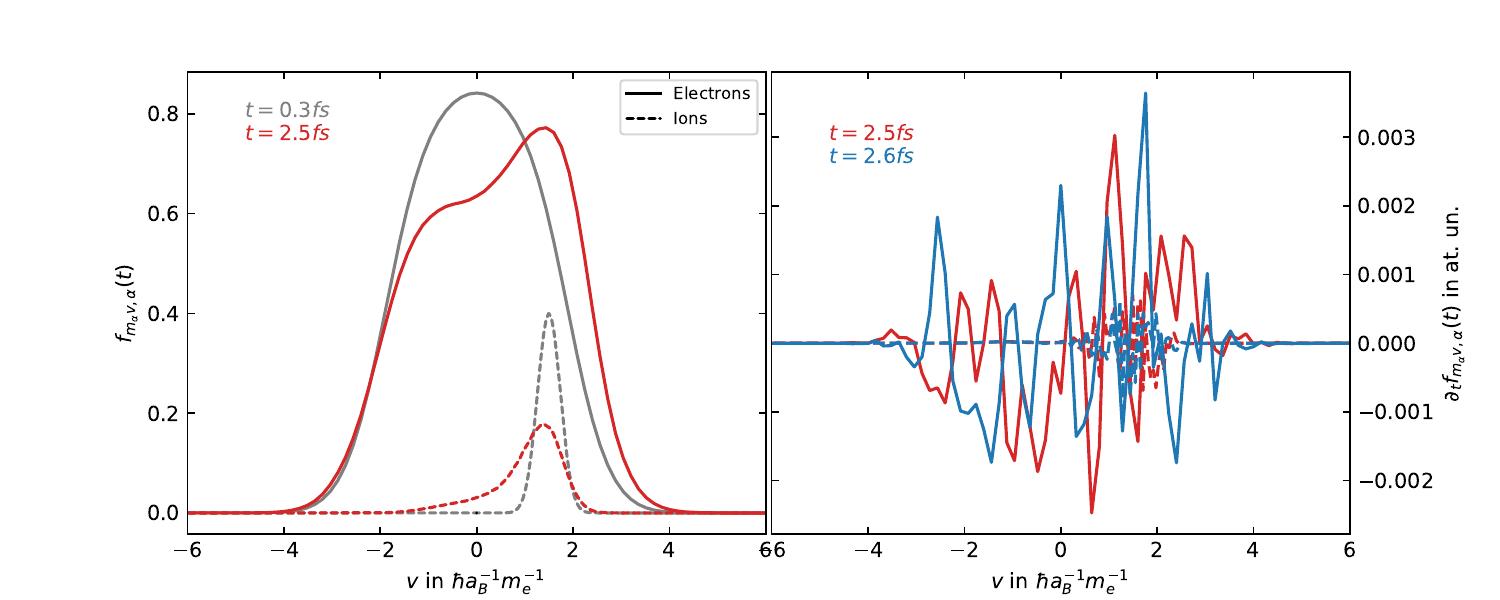}
    \caption{First signs of an aliasing problem. Left: velocity distribution functions of electrons and ions in a uniform 1D system. Electronic parameters: Fermi distribution with $r_s=0.5$, $\beta=1.0\,Ha^{-1}$. Ion parameters: $m_i=3m_e$, Gaussian distribution centered around $p=4.5\hbar a_B^{-1}$ with height $0.4$ and $\sigma^2=0.5\hbar^2a_B^{-2}$. The $k$-point spacing is given by $\Delta k=0.16\hbar a_B^{-1}.$ The initial distributions (after adiabatic switching of $e-e$ and $i-i$ correlations) are depicted by the grey lines which then relax due to $e-i$ collisions. The equilibration seems to stop after a small period of time in a state that does not resemble a Fermi distribution. Right: The time-derivative of the distribution function reveals a complicated and volatile peak structure, which is due to aliasing. Since the peaks quickly shift and change sign within short time intervals, the net relaxation is artificially slowed down. With the present diffusion approach, these artifacts are removed, cf. Fig.~\ref{fig:TDDist}.}
    \label{fig:DemoAlias}
\end{figure}
The G1-G2 scheme has been extensively solved numerically for lattice models, e.g. Refs.~\onlinecite{joost_prb_20,joost_prb_22}. For spatially uniform systems solutions in the momentum representation (continuous basis) are severly hampered by the very large basis dimension. For this reason, so far, only quasi-1D systems could be studied with the G1-G2 scheme\cite{makait_cpp_23}. These simulations, for long time propagations became unstable, as is demonstrated in Fig.~\ref{fig:DemoAlias}: the time derivative of the momentum distribution develops artificial oscillations, see right part of Fig.~\ref{fig:DemoAlias}.  
The origin of this behavior is the finite discretization of the momentum space and can be understood from the SOA collision integral\footnote{\footnotesize In principle, a sufficiently  accurate numerical solution of this integral from known values of $G^\gtrless$ is possible: using interpolations of the second line of Eq.~\eqref{eq:CollIntSOAMemory}, one can artificially increase the resolution of the grid and thereby avoid aliasing. However, such a procedure is expensive, and an extension to more advanced selfenergies turns out much more difficult and practically unfeasible.}, Eq.~\eqref{eq:CollIntSOAMemory}.
The 1-time NEGF are purely imaginary, hence the term in the second line is purely real and typically varies slowly as a function of $\mathbf{p},\mathbf{p}_2,$ and $\mathbf{q}$. The exponential,  as a function of $\mathbf{p},\mathbf{p}_2,\mathbf{q}$, produces increasingly dense oscillations as $\tau$ increases. In simulation programs, the Green functions are stored and computed on a grid. Assuming a cartesian grid with spacing $\Delta k$, integrals are treated using the quadrature 
\begin{align}
    \int d\mathbf{k}\,f(\mathbf{k})\longrightarrow\frac{1}{(\Delta k)^d}\sum\limits_{\mathbf{k}}f_\mathbf{k}\,.
\end{align}
For $\tau$, in the memory integral, sufficiently large, the phase difference between neighboring grid points will also become large. If the oscillations are more dense than the momentum grid can resolve, the integral computed from the discretization is expected to be faulty. This is a form of aliasing, as the integral is the $0$-component of the Fourier transform. The same phenomenon appears in the G1--G2 scheme in the behavior of $\mathcal{G}$, cf. Eq. \eqref{eq:col-int-g2}. In SOA, we have 
\begin{align}
    \mathcal{G}_{\mathbf{kpq}}^{\alpha\beta}(t)=\frac{1}{i\hbar}\int_{t_0}^t d\bar t\,\Psi_{\mathbf{kpq}}^{\alpha\beta}(\bar{t}) \exp{\left(\frac{-i}{\hbar }\int_{\bar{t}}^t\left[h^{\text{HF}}_{\mathbf{k}+\mathbf{q},\alpha}+h^{\text{HF}}_{\mathbf{p}-\mathbf{q},\beta}-h^{\text{HF}}_{\mathbf{k},\alpha}-h^{\text{HF}}_{\mathbf{p},\beta}\right]_{t'}dt'\right)}\,,
\end{align}
with $h^\mathrm{HF}$ evaluated at $t'$, and  $\Psi^{\alpha\beta}$ was given by Eq.~\eqref{eq:defPsi}.

The problem of the increasingly rapid oscillations that causes the aliasing arises from the momentum-dependent phase, $\omega_{\mathbf{kpq}}^{\alpha\beta}\tau \equiv \left[h^{\text{HF}}_{\mathbf{k}+\mathbf{q},\alpha}+h^{\text{HF}}_{\mathbf{p}-\mathbf{q},\beta}-h^{\text{HF}}_{\mathbf{k},\alpha}-h^{\text{HF}}_{\mathbf{p},\beta}\right][t-\bar t]/\hbar$, which involves combinations of single-particle energies 
\begin{align}
    \omega_{\mathbf{kpq}}^{\alpha\beta} &= \omega_{\mathbf{kq}}^{\alpha} + \omega_{\mathbf{p,-q}}^{\beta}\,,\label{eq:omega-kpq}\\
    \hbar \omega_{\mathbf{kq}}^{\alpha} &= h^{\text{HF}}_{\mathbf{k}+\textbf{q},\alpha} - h^{\text{HF}}_{\mathbf{k},\alpha} \,.\label{eq:omega-kq}
\end{align}
To derive and illustrate our regulazation procedure we simplify the analysis by neglecting the time dependence of the Hartree-Fock energies. If necessary, this dependence can be restored later.

The phase difference between neighboring grid points in an arbitrary $k_m$-direction can be approximated by
\begin{align}
    \Delta\varphi = \tau \left|\omega_{\mathbf{kpq}}^{\alpha\beta} - \omega_{\mathbf{k} +\Delta k\hat{e}_m,\mathbf{pq}}^{\alpha\beta}\right| \approx \tau \Delta k \left|\frac{\partial \omega}{\partial k_m}\right|\,,\label{eq:phi}
\end{align}
where large $\Delta\varphi$ appear at different rates in different directions. Our aim will be to suppress the emerging aliasing before it becomes too strong. Therefore, we define the ``aliasing times'', $\tau^k_\mathrm{alias}$ and $\tau^p_\mathrm{alias}$, after which a regularization will be performed that suppresses artificial instabilities on the momentum lattice in $p$- and $k$-directions, respectively. For this we choose a threshold value for the phase difference, Eq.~\eqref{eq:phi}, $\Delta \phi = 1$, which yields
\begin{align}
    \tau^k_\mathrm{alias}=\frac{1}{\Delta k |\nabla_{\mathbf{k}} \omega|}\qquad,\qquad \tau^p_\mathrm{alias}=\frac{1}{\Delta k |\nabla_{\mathbf{p}} \omega|}\,.\label{eq:tau-alias}
\end{align}
Note that, in general, these are still functions of $\mathbf{k,p,q},\alpha$ and $\beta$, implying that aliasing does not emerge at the same rate and time in all momentum space regions. 

In the free GKBA (F-GKBA), the behavior can be understood more easily. Then the two-particle energy difference, Eq.~\eqref{eq:omega-kpq}, becomes 
$\hbar\omega_{\mathbf{kpq}}^{\alpha\beta}=\left[E^\alpha(\textbf{k}+\textbf{q})+E^\beta(\textbf{p}-\textbf{q}) - E^\alpha(\textbf{k})-E^\beta(\textbf{p})  \right]$, and the gradient, $\nabla_{\mathbf{kp}}\equiv (\nabla_{\mathbf{k}},\nabla_{\mathbf{p}})$, becomes
$\nabla_{\mathbf{kp}}\omega_{\mathbf{kpq}}^{\alpha\beta}=\left(\frac{1}{m_\alpha}\mathbf{q},-\frac{1}{m_\beta}\mathbf{q}\right)$.

\subsection{Diffusion approach in the G1--G2 scheme}\label{sec:TheoryDiffusionG1G2}
The large phase differences on the momentum grid, Eq.~\eqref{eq:phi}, become critical when $\tau$ increases, i.e., in case of strong memory effects. At the same time, we expect that the memory depth will always be finite, being on the order of the correlation time \cite{bonitz96pla}, due to higher order interactions (that are neglected in the HF-GKBA) or dephasing effects which would lead to a damping of these oscillations and remove the aliasing problem. Such a damping occurs naturally in the full two-time NEGF theory. Within the GKBA, one can include damping phenomenologically via the \textit{Lorentzian Hartree--Fock} GKBA (LHF-GKBA)\cite{bonitz_98teubner}
\begin{align}
    h^\text{LHF}_{\mathbf{p}\sigma}(t)=h^\text{HF}_{\mathbf{p}\sigma}(t) + \mathrm{i}\hbar\gamma\,.
    \label{eq:hhf-damped}
\end{align}
However, it is well known that the ansatz \eqref{eq:hhf-damped}  violates total energy conservation. One approach to improve the LHF-GKBA is to replace the exponential time dependence by $\cosh$-functions in the GKBA, as was proposed in Ref.~\onlinecite{bonitz-etal.99epjb}. In that reference it was found that this ansatz improves the short-time (large frequency) limit of the spectral function, however, it only partially improves the energy conservation. Therefore, even though these damped modifications of the HF-GKBA remove the aliasing, here, we will be following a different path.

In this paper we develop a new approach that reduces aliasing, preserves energy conservation and, in addition, leads only to minimal modifications of the equations of motion of the G1-G2 scheme. Our starting point is the relation of the correlation energy of a uniform gas to the two-particle correlation function:
\begin{align}
    E_\mathrm{corr}(t)&=-\frac{1}{2}\mathfrak{Im}\left\{\sum\limits_{\mathbf{kpq}\alpha\beta} \mathcal{G}_{\mathbf{kpq}}^{\alpha\beta}(t)w_\mathbf{q}\right\}\,.
\end{align}
This implies that any transformation of $\mathcal{G} \to \tilde{\mathcal{G}}$ that preserves the sums 
\begin{align}
    \sum\limits_{\textbf{kp}}\mathcal{G}_{\textbf{kpq}}^{\alpha\beta}\equiv \sum\limits_{\textbf{kp}}\Tilde{\mathcal{G}}_{\textbf{kpq}}^{\alpha\beta}
\end{align}
also conserves the correlation energy\footnote{An extension to more dependencies of $w_{\mathbf{kpq}}$, as they appear in multi-band systems in solids, can be achieved analogously by carrying some additional balancing weights through the derivation.}. 
This finding suggests that an energy-conserving way to mitigate aliasing can be formulated in terms of such a transformation. A rather general class of such transformations is given by convolutions with functions $M_\mathbf{k}$ which fulfill
\begin{align}\label{eq:g2-trafo}
    \Tilde{\mathcal{G}}^{\alpha\beta}_{\mathbf{kpq}}&=\sum\limits_{\tilde{\mathbf{k}},\tilde{\mathbf{p}}} M_{\mathbf{k}-\tilde{\mathbf{k}}} M_{\mathbf{p}-\tilde{\mathbf{p}}} \mathcal{G}^{\alpha\beta}_{\tilde{\mathbf{k}}\tilde{\mathbf{p}}\mathbf{q}}\,,\qquad\mbox{with}\quad
    \sum\limits_\mathbf{k} M_\mathbf{k}=1\,.
\end{align}
A particularly simple transformation kernel is given by
\begin{align}
    M_0&=1-2d\alpha \\
    M_{\pm \Delta_k \hat{e}_i}&=\alpha\\
    M_\mathbf{k}&=0\,,\quad\text{else}\,,
\label{eq:laplace-trafo}
\end{align}
i.e., we have nonzero values on the $d$-dimensional cross through $\mathbf{k}=0$, extending $\Delta_k$ along every coordinate axis. $\alpha$ is a real parameter. If $\alpha>0$ it leads to a reduction of the amplitude of dense oscillations. This is no coincidence since this choice of $M$ has a close relation to the discrete Laplacian operator,
\begin{align}
    M_\mathbf{k}=\delta_{\mathbf{k}0} + \alpha\Delta^\mathrm{discrete}_\mathbf{k}\,,\qquad \Delta_\mathbf{k}^\mathrm{discrete}=\begin{cases}
        -2d/(\Delta k)^2,\qquad &\mathbf{k}=0\\
        +1/(\Delta k)^2, &\mathbf{k}=\pm \Delta_k \hat{e}_i\quad\text{for an }i\in\{1,..,d\}\\
        0, & \text{else }
    \end{cases}
\end{align}
The spectrum of the Laplacian operator is given by $\Delta_\mathbf{k} e^{i\mathbf{r}\cdot\mathbf{k}}=-r^2e^{i\mathbf{r}\cdot\mathbf{k}}.$ As in the case of the diffusion equation, such a Laplacian reduces the amplitude of oscillations over time. Only if $\alpha$ is too large, one could experience overshooting, and a reduction of the amplitude does not occur. Since we are discussing oscillations on a momentum/wavenumber grid, the corresponding 'frequencies' are given by (relative) positions. 

There are many more conserving transformations, of which some are relevant here: Let $M_{\mathbf{k}\tilde{\mathbf{k}}}$ be a matrix with $\sum_\mathbf{k}M_{\mathbf{k}{\tilde{\mathbf{k}}}}=0$ such as $M=\Delta^\mathrm{discrete}.$ Then for any grid function $D$,
\begin{align}
    \mathcal{M}_{\mathbf{kp},\tilde{\mathbf{k}}\tilde{\mathbf{p}}}^\mathbf{q} &= \delta_{\mathbf{k}\tilde{\mathbf{k}}}\delta_{\mathbf{p}\tilde{\mathbf{p}}} + D_{\tilde{\mathbf{k}}\tilde{\mathbf{p}}\mathbf{q}}^{k}M_{\mathbf{k}\tilde{\mathbf{k}}}+D_{\tilde{\mathbf{k}}\tilde{\mathbf{p}}\mathbf{q}}^{p}M_{\mathbf{p}\tilde{\mathbf{p}}}
\end{align}
defines a conserving transformation by
\begin{align}
    \Tilde{\mathcal{G}}^{\alpha\beta}_{\mathbf{kpq}}&=\sum\limits_{\tilde{\mathbf{k}},\tilde{\mathbf{p}}} \mathcal{M}_{\mathbf{kp},\tilde{\mathbf{k}}\tilde{\mathbf{p}}}^\mathbf{q} \mathcal{G}^{\alpha\beta}_{\tilde{\mathbf{k}}\tilde{\mathbf{p}}\mathbf{q}}
\end{align}
Clearly, this is of matrix-vector product form. Choices, such as $D\sim \omega$ or $D\sim |\nabla\omega|$, that will be considered below are in this class and thus conserving. Having found conserving transformations the next questions concern how often and when to apply these transformations. Since aliasing emerges at different rates in different parts of the phase space, and the emergence rate is unbounded on the whole phase space (cf. Eq. \eqref{eq:tau-alias}) it is purposeful to apply them as often as possible (to counteract in time) but with rescaled amplitudes $D$ to keep the action finite:
\begin{align}
    \Tilde{\mathcal{G}}&=\left(1 + \frac{D^k}{n}\Delta^{\mathrm{discrete},k}+ \frac{D^p}{n}\Delta^{\mathrm{discrete},p}\right)^n\mathcal{G}\overset{n\to\infty}{\longrightarrow}\exp\left(\left(D^k\Delta^{\mathrm{discrete},k}+D^p\Delta^{\mathrm{discrete},p}\right)t\right)\mathcal{G}\,,
\end{align}
with $t=1$. Since the expression is conserving for any finite $n$, the limit is as well. This transformation is also the solution of a diffusion equation which suppresses fine structures of $\mathcal{G}$ such as critically dense oscillations.
The regularization procedure now is in a form that can be added to the $GW$ equation of motion, giving 
\begin{align}
    \mathrm{i}\hbar \frac{\text{d}}{\text{d}t}\tilde{\mathcal{G}}^{\alpha\beta}_{\mathbf{kpq}}(t)-\left[h^{\text{HF},(2)}(t),\tilde{\mathcal{G}}(t)\right]_{\mathbf{kpq}}^{\alpha\beta}&=\Psi^{\alpha\beta}_{\mathbf{kpq}}(t)+\Pi^{\alpha\beta}_{\mathbf{kpq}}(t)+i\hbar \left(D_\mathbf{kpq}^{k,\alpha\beta}\Delta_\mathbf{k} + D_\mathbf{kpq}^{p,\alpha\beta}\Delta_\mathbf{p}\right)\tilde{\mathcal{G}}\label{eq:GWG2Diff1}\,,
\end{align}
where we introduced two positive momentum-space diffusion constants, $D^{k,\alpha\beta}_\mathbf{kpq}$ and $D^{p,\alpha\beta}_\mathbf{kpq}$, that can (but do not need to) depend on the particle species/spin and the momenta. In the following, we will just write $\mathcal
{G}$ instead of $\tilde{\mathcal{G}}$. From dimensionality considerations, the momentum diffusion coefficient is expected to be proportional to the square of the momentum grid length and inversely proportional to the relevant time step for which it is reasonable to choose the value \eqref{eq:tau-alias}:
\begin{align}
    D^{k,\alpha\beta}_\mathbf{kpq}= 3\Gamma \frac{(\Delta k)^2}{\tau^k_\mathrm{alias}}=3\Gamma (\Delta k)^3 \left|\nabla_\mathbf{k} \omega_{\mathbf{kpq}}^{\alpha\beta}\right|\,,\qquad\qquad D^{p,\alpha\beta}_\mathbf{kpq}= 3\Gamma \frac{(\Delta k)^2}{\tau_\mathrm{alias}^{p}}=3\Gamma (\Delta k)^3 \left|\nabla_\mathbf{p} \omega_{\mathbf{kpq}}^{\alpha\beta}\right|\,.
\end{align}
Here, $\Gamma$ is a dimensionless free parameter the choice of which allows one to control the rate at which aliasing is removed. While large $\Gamma$ suppresses aliasing more effectively, this also represents a stronger intervention into the dynamics -- hence a compromise has to be made. The dependence of $D$ on $\tau_\mathrm{alias}^{-1}$ makes it effective in acting stronger at such momentum combinations where the aliasing emerges fastest.

In Sec.~\ref{sec:3} we will demonstrate that this approach, indeed, removes aliasing. We will present numerical results for different values of $\Gamma$ and analyze the sensitivity of the results to this choice. But before turning to simulations, we present an independent derivation of the diffusion equation \eqref{eq:GWG2Diff1} via a k-space average of the G1--G2 equations.\footnote{
Another physical motivation for the diffusion approach and choice of parameters can be obtained from the non-Markovian expressions of the collision integrals and will be given in a forthcoming publication, cf. Ref.~\onlinecite{makait_pop_25}
}

\subsection{Motivation of the Diffusion approach by  k-space averaging}\label{sec:DiffusionG1G2-average}
Denoting the k-space average by $\langle \dots \rangle$, an arbitrary function in momentum space can be written as 
$A_k=\langle A\rangle_k + \delta A_k$.
Following Ref.~\onlinecite{bonitz_psst18}, we now average Eq.~\eqref{eq:G2GW}, where we skip the species indices and the time arguments:
\begin{align}
    \mathrm{i}\hbar \frac{\text{d}}{\text{d}t}\langle\mathcal{G}\rangle_{\mathbf{kpq}} = & 
    \langle\mathcal{G}\rangle_{\mathbf{kpq}}\langle\hbar\omega\rangle_{\mathbf{kpq}}
    +\langle \Psi^{\pm}\rangle_{\mathbf{kpq}}+\langle\Pi\rangle_{\mathbf{kpq}} 
    + \langle \delta\mathcal{G}_{\mathbf{kpq}}\delta \hbar\omega_{\mathbf{kpq}}\rangle\,, \label{eq:g2eq-average1}
\end{align}
and the last term arises from averaging the product of two random functions.
Usually one proceeds by expressing all terms via the average quantities in order to eliminate irrelevant or unphysical small-scale fluctuations where the last term acts as an additional contribution (``collision integral'') that contains the trace of the small scales \cite{bonitz_psst18}. Here we proceed differently and express the average via the exact function value at a given point, weighted by values in its neighborhood.
Starting with the simplest possible (lowest order) k-space average in one dimension, we only include the two adjacent points on the grid, i.e.
\begin{align}\label{eq:average-def}
    \langle A\rangle_k \equiv \frac{1}{2} A_k + \frac{1}{4}(A_{k+\Delta k}+A_{k-\Delta k})\,,
\end{align}
which leads to 
\begin{align}\label{eq:delta-a}
    \delta A_k = \frac{1}{2} A_k - \frac{1}{4}(A_{k+\Delta k}+A_{k-\Delta k})\,.
\end{align}
Assuming that the grid is fine enough, i.e. $\Delta k$ is sufficiently small, we can Taylor expand the terms in the parantheses around $k$: 
\begin{align}
    \frac{1}{4}(A_{k+\Delta k}+A_{k-\Delta k}) = \frac{1}{2}A_k + \frac{1}{4}A_k''(\Delta k)^2 + \mathcal{O}[(\Delta k)^4]\,,
\end{align}
which yields, for the average and for the fluctuation,
\begin{align}\label{eq:mean-a}
\langle A\rangle_k &= A_k + \frac{1}{4}A_k''(\Delta k)^2 + \mathcal{O}[(\Delta k)^4]\,,\\
\delta A_k &= - \frac{1}{4}A_k''(\Delta k)^2 + \mathcal{O}[(\Delta k)^4]\,.\label{eq:delta-a}
\end{align}

Applying these results to the first term on the r.h.s. of Eq.~\eqref{eq:g2eq-average1}, we obtain
\begin{align}\label{eq:product-averages2}    &\langle\mathcal{G}\rangle_{\mathbf{kpq}}\langle\hbar\omega\rangle_{\mathbf{kpq}} = \mathcal{G}_{\mathbf{kpq}}\hbar\omega_{\mathbf{kpq}} + \\
    &+\frac{(\Delta \Vec{k})^2}{4}\left\{\hbar\omega_{\mathbf{kpq}}\Delta_\textbf{k}\mathcal{G}_{\mathbf{kpq}} + \mathcal{G}_{\mathbf{kpq}}\Delta_\textbf{k}\left( U^{\rm HF}_{\textbf{k}-\mathbf{q}} - U^{\rm HF}_\textbf{k}\right)\right\} + 
    \frac{(\Delta \Vec{p})^2}{4}\left\{\hbar\omega_{\mathbf{kpq}}\Delta_\textbf{p}\mathcal{G}_{\mathbf{kpq}} + \mathcal{G}_{\mathbf{kpq}}\Delta_\textbf{p}\left( U^{\rm HF}_{\textbf{p}+\mathbf{q}} - U^{\rm HF}_\textbf{p}\right)\right\} \,,
\end{align}
where we took into account that, in case of a parabolic dispersion, the free-particle energies do not contribute to the second derivative of $\omega_{\mathbf{kpq}}$, because $E''_\textbf{k}=E''_{\textbf{k}-\mathbf{q}}=\frac{1}{m}$, and the difference of the two vanishes. We now assume that the grid spacings in k- and p-directions are the same, $\Delta p=\Delta k$, and insert the result \eqref{eq:product-averages2} into Eq.~\eqref{eq:g2eq-average1}:
\begin{align}
    \mathrm{i}\hbar \frac{\text{d}}{\text{d}t}\langle\mathcal{G}\rangle_{\mathbf{kpq}} \approx &     \mathcal{G}_{\mathbf{kpq}}\hbar\omega_{\mathbf{kpq}}
    + \langle \Psi^{\pm}\rangle_{\mathbf{kpq}}+\langle\Pi\rangle_{\mathbf{kpq}} \label{eq:g2eq-average-final}
    + \frac{(\Delta \Vec{k})^2}{4}\hbar\omega_{\mathbf{kpq}}(\Delta_\textbf{k}+\Delta_\textbf{p})\mathcal{G}_{\mathbf{kpq}} + \\
        &+ \frac{(\Delta \Vec{k})^2}{4}\hbar\omega_{\mathbf{kpq}}\mathcal{G}_{\mathbf{kpq}}\left\{\Delta_\textbf{k}(U^{\rm HF}_{\textbf{k}-\mathbf{q}} - U^{\rm HF}_\textbf{k}) +
         \Delta_\textbf{p}(U^{\rm HF}_{\textbf{p}+\mathbf{q}} - U^{\rm HF}_\textbf{p} )\right\} + \mathcal{O}(\Delta k)^3\,,
\end{align}
where the last term in Eq.~\eqref{eq:g2eq-average1} has been dropped since it is of order $(\Delta \Vec{k})^4$. In many cases the final two lines will be small since they are of the order of the third derivative of the Hartree-Fock energies with respect to k or p and scale as $(\Delta k)^2 q$ [due to the long-range nature of the pair interaction we expect that small-$q$ contributions will dominate]. Note also that we use, approximately $\langle \Psi^{\pm}\rangle(G) \approx \Psi^\pm(\langle G \rangle)$ and $\langle \Pi\rangle(G,\mathcal{G}) \approx \Pi(\langle G \rangle, \langle\mathcal{G}\rangle)$, neglecting terms of the form $\langle \delta G \delta G\rangle$ and $\langle \delta G \delta \mathcal{G}\rangle$, as they will be of order $(\Delta k)^4$.
If the system (\ref{eq:g2eq-average-final}) is propagated in time, the function value for the next time step, $\langle\mathcal{G}\rangle_{\mathbf{kpq}}(t+\Delta t)$  is, in lowest order, expressed via $\langle\mathcal{G}\rangle_{\mathbf{kpq}}(t) + \Delta t \cdot \mathcal{F}[\mathcal{G}_{\mathbf{kpq}}(t)\to \langle\mathcal{G}\rangle_{\mathbf{kpq}}(t)]$, where $\mathcal{F}$ denotes the r.h.s. and, in its arguments, the original function $\mathcal{G}$ is replaced by the average $\langle\mathcal{G}\rangle$. This means, at each time step, the averaging procedure is carried out which smoothens fluctuations of $\mathcal{G}$ and $\omega$ between neighboring grid points\footnote{In this sense, the procedure is similar to purification, cf. Refs.~\onlinecite{lackner_PhysRevA.91.023412, joost_prb_22,donsa_prr_23}. But instead of removing unstable eigenvalues from the two-particle eigenvalue spectrum, here, unstable $k$-modes are removed from the Fourier spectrum.}.

Let us now compare the result (\ref{eq:g2eq-average-final}) to the diffusion equation, Eq.~\eqref{eq:GWG2Diff1} that was obtained in Sec.~\ref{sec:TheoryDiffusionG1G2}. The terms in   Eq.~\eqref{eq:GWG2Diff1} coincide with the corresponding expressions in Eq. \eqref{eq:g2eq-average-final}, if the (small) contributions from the third derivatives of $U^\mathrm{HF}$ are neglected. 
This means that the lhs of Eq.~\eqref{eq:GWG2Diff1} agrees with the lhs of Eq.~\eqref{eq:g2eq-average-final}, i.e., with the k-space average of the time derivative, confirming the correctness of our interpretation of Eq.~\eqref{eq:GWG2Diff1}. 
Finally, the terms with the Laplace operators have the same structure, and they would exactly coincide if the diffusion coefficients would be chosen according to
\begin{align}
    i\hbar D_\mathbf{kpq} = 3 i\hbar\Gamma \frac{(\Delta k)^2}{\tau_\mathrm{alias}^{p}} \equiv
    \frac{(\Delta \Vec{k})^2}{4}\hbar\omega_{\mathbf{kpq}}\,.
\end{align}
However, we are not primarily interested in a momentum average (coarse graining) of the $\mathcal{G}$-equation. Instead, it is sufficient to suppress instabilities in $k-$space where they are most critical and leave $\mathcal{G}$ in the remaining parts of the momentum space un-altered. As we will demonstrate below, the choice of $\tau_{\rm alias}$ and $\Gamma$ that was discussed in Sec.~\ref{sec:TheoryDiffusionG1G2} is very sufficient in mitigating aliasing.

\section{Simulation Results}\label{sec:3}

Aliasing is particularly strong in simulations of 1D systems, due to the limited number of different possible scattering events. Therefore, below we present quasi-1D simulations. 

But before doing this we present an analytical illustration of aliasing in 1D.
To this end we neglect, for a moment, the time-dependence of the single-particle Green functions in Eq.~\eqref{eq:CollIntSOAMemory}, and perform the time integration. Considering only the real part (relevant for changes of $f$), and abbreviating the scattering weights in the second line of Eq.~\eqref{eq:CollIntSOAMemory} by $\Phi_{\mathbf{p}\mathbf{p}_2\mathbf{q}}^{\alpha\beta}(t)$, we can write
\begin{align}
    \mathfrak{Re}I_{\mathbf{p}\alpha}(t)&=(i\hbar)^2\sum\limits_{\beta} \int_{0}^{t-t_0}\text{d}\tau\int\frac{\text{d}\mathbf{p}_2}{(2\pi\hbar)^d}\int\frac{\text{d}\mathbf{q}}{(2\pi\hbar)^d}w_{\mathbf{q}}\left[w_\mathbf{q}\pm \delta_{\alpha\beta}w_{\mathbf{p}-\mathbf{p}_2-\mathbf{q}}\right]\cos\left[\omega_{\mathbf{p}\mathbf{p}_2\mathbf{q}}^{\alpha\beta}\tau\right]\Phi_{\mathbf{p}\mathbf{p}_2\mathbf{q}}^{\alpha\beta}(t)\\
    &=(i\hbar)^2\sum\limits_{\beta} \int\frac{\text{d}\mathbf{p}_2}{(2\pi\hbar)^d}\int\frac{\text{d}\mathbf{q}}{(2\pi\hbar)^d}w_{\mathbf{q}}\left[w_\mathbf{q}\pm \delta_{\alpha\beta}w_{\mathbf{p}-\mathbf{p}_2-\mathbf{q}}\right]\frac{\sin\left[\omega_{\mathbf{p}\mathbf{p}_2\mathbf{q}}^{\alpha\beta}(t-t_0)\right]}{\omega_{\mathbf{p}\mathbf{p}_2\mathbf{q}}^{\alpha\beta}}\Phi_{\mathbf{p}\mathbf{p}_2\mathbf{q}}^{\alpha\beta}(t)\,.
\end{align}
This approximation is known as collisional broadening approximation \cite{bonitz_qkt} which approaches the Markovian Landau collision integral, for long times.
With growing $\tau$, the ratio $\frac{\sin\left[\omega_{\mathbf{p}\mathbf{p}_2\mathbf{q}}^{\alpha\beta}(t-t_0)\right]}{\omega_{\mathbf{p}\mathbf{p}_2\mathbf{q}}^{\alpha\beta}}$ becomes increasingly sharp (in a distributional sense) around points $(\mathbf{p},\mathbf{p}_2,\mathbf{q})$ where $\omega_{\mathbf{p}\mathbf{p}_2\mathbf{q}}^{\alpha\beta}=0$. 
The resonance condition can be written for the F-GKBA as $0 = \mathbf{q}\cdot (\mathbf{v}-\mathbf{v}_2 + (\mathbf{q}/2m_\alpha + \mathbf{q}/2m_\beta))$. If small $\mathbf{q}$ dominate (Coulomb divergence), this condition yields $\mathbf{q}\perp(\mathbf{v}-\mathbf{v}_2)$. While there are a variety of momentum combinations that can fulfill this in 2D and 3D which leads to the possibility of many different scattering processes happening, this is very restricted in 1D as it implies $\mathbf{v}\approx \mathbf{v}_2.$ 
On a finite grid, this condition leads to a poorly resolved function ${\sin\left[\omega_{\mathbf{p}\mathbf{p}_2\mathbf{q}}^{\alpha\beta}(t-t_0)\right]}/{\omega_{\mathbf{p}\mathbf{p}_2\mathbf{q}}^{\alpha\beta}}$, giving rise to erroneous collision integrals. The consequence are unphysical fluctuating `spikes' in the momentum and time dependence of $\frac{df}{dt}$ that are clearly visible in Figs.~\ref{fig:DemoAlias} and \ref{fig:GWDynDeriv}.
 
\subsection{Quasi-stationary SOA in 1D}\label{ssec:QStatSOA}
\begin{figure}[h]
    \centering
    \includegraphics[width=0.8\linewidth]{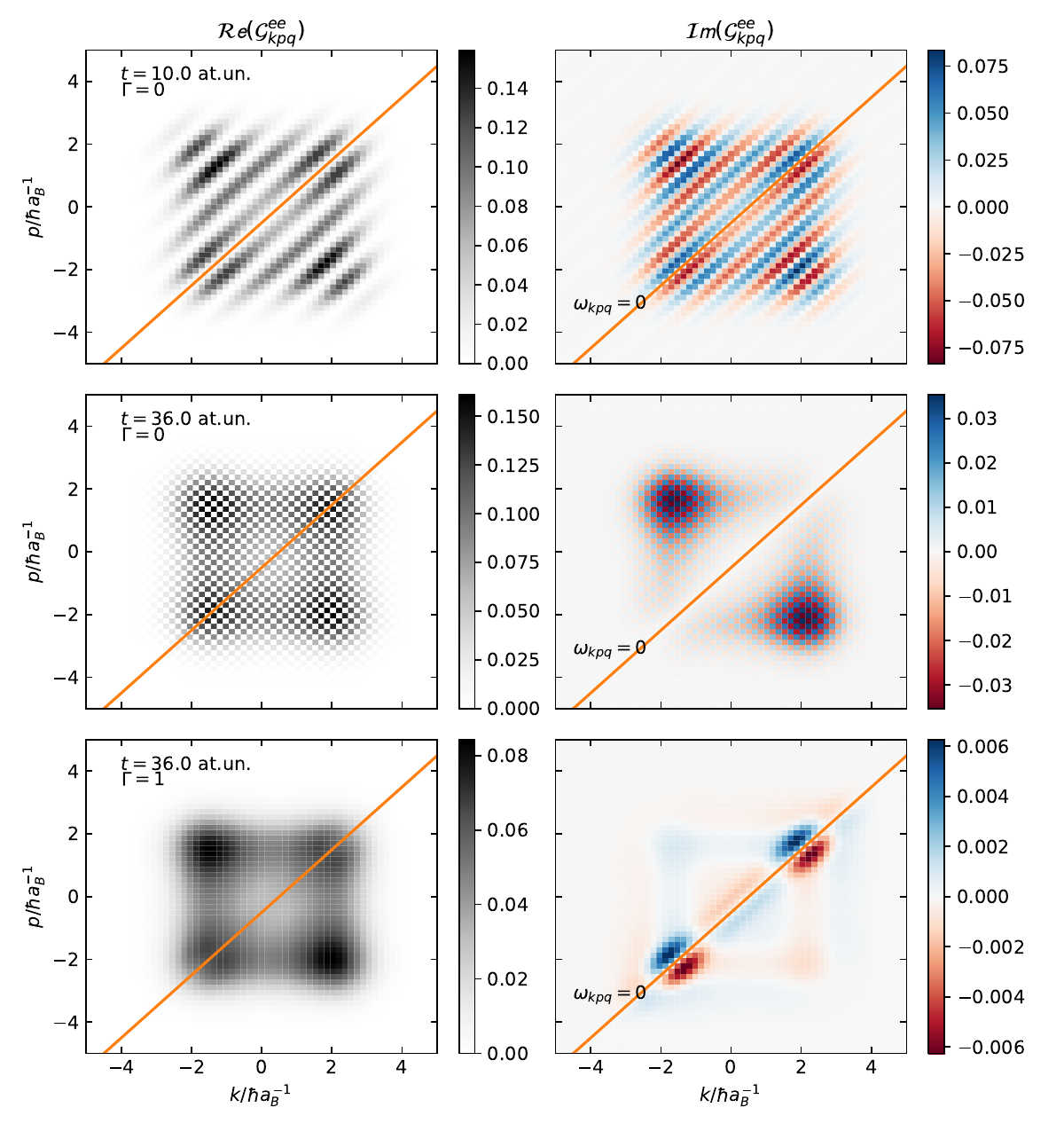}
    \caption{Momentum dependence of $\mathcal{G}_{\mathbf{kpq}}^{e\uparrow,e\downarrow}(t)$ in a uniform 1D plasma with a time-independent electron distribution, for  $q=0.51\hbar a_B^{-1}$. Top two rows: no diffusion, $\Gamma=0$. 
    Left (right) column: real (imaginary) part. Top: early time of $t=10$a.u., two bottom rows: later time of $t=36$a.u. 
    Each square represents one grid point in the simulation. For the later time, the grid is insufficient to resolve the oscillations giving rise to aliasing. Bottom row: same as middle row, but with diffusion, $\Gamma=1$ which removes the artificial oscillations. Instead, only a few spots near the $\omega_{kpq}=0$ line are notably occupied, that corresponds to Fermi's golden rule. The system state at $t=10$  is indistinguishable from the one for $\Gamma=0$, shown in Fig. \ref{fig:G2_t50_G0}. The symmetry is a sign of an equilibrium state (here Fermi distribution).
    The positive real part of $\mathcal{G}$ corresponds to a negative real part of the two-particle density matrix, and thus implies a negative correlation energy. The imaginary part is relevant for changes in the distribution function, i.e. in $G^\gtrless$. 
    }
    \label{fig:G2_t50_G0}
\end{figure}

We discuss the properties of our approach, in particular, the influence of the diffusion term, using the example of particle stopping in dense 1D-plasmas, cf. Ref.~\onlinecite{makait_cpp_23} for details about the model. In this section we first analyze the novel approach in a simplified situation, where we treat correlations on SOA level and neglect the time-dependence of $G^\gtrless$ and use the free single-particle energies (F-GKBA), since, in this limit we can understand the effects analytically. Fig. \ref{fig:TDDist} (0.3fs) shows what this frozen distribution looks like. All results in this section were computed with a discretization of $\Delta k= \frac{1}{6}\hbar a_B^{-1}$. The time-dependent results are discussed in the next subsection.

Fig. \ref{fig:G2_t50_G0} demonstrates the emergence of aliasing directly on the level of $\mathcal{G}$. It also explains how the diffusion term removes it entirely. If $G^\gtrless$ is time-independent, the $\mathcal{G}$-equation becomes linear and inhomogeneous, and the solution has the form
\begin{align}
    \mathcal{G}_{\mathbf{kpq}}^{\alpha\beta}(t)=\mathcal{G}_{\mathbf{kpq}}^{\mathrm{stat},\alpha\beta}+\mathcal{G}_{\mathbf{kpq}}^{\mathrm{amp},\alpha\beta}e^{-i\omega_{\mathbf{kpq}}^{\alpha\beta} t}\,,
\end{align}
where $\mathcal{G}^{\mathrm{stat}}$ is the time-independent solution of the $\mathcal{G}$-equation, whereas $\mathcal{G}^\mathrm{amp}$ denotes the difference from the initial value
\begin{align}
    \mathcal{G}^\mathrm{stat}=\frac{\Psi}{\hbar\omega}\,,\qquad \mathcal{G}^\mathrm{amp}=\mathcal{G}(0)-\mathcal{G}^\mathrm{stat}\,.
\end{align}
Geometrically, this means that, for fixed $\mathbf{kpq}$, $\mathcal{G}(t)$ draws a circle of radius $\mathcal{G}_{\mathbf{kpq}}^{\mathrm{amp},\alpha\beta}$ around the center, $\mathcal{G}_{\mathbf{kpq}}^{\mathrm{stat},\alpha\beta}$, with the frequency $\omega_{\mathbf{kpq}}^{\alpha\beta}$. Since $\omega_{\mathbf{kpq}}^{\alpha\beta}$ is linear (in free GKBA) in $\mathbf{k}$ and $\mathbf{p}$, this yields a plane wave pattern in the real and imaginary parts of $\mathcal{G}$, which can be seen in Fig.~\ref{fig:G2_t50_G0}. It can be shown analytically\cite{bonitz_qkt} that this pattern eventually approaches a $\delta(\omega_\mathbf{kpq}^{\alpha\beta})$-function in a distributional sense, in agreement with Fermi's golden rule. Numerically, we cannot resolve $\mathcal{G}$ anymore once the oscillations in the plane wave pattern become too dense, cv. middle row of Fig.~\ref{fig:G2_t50_G0}, what prevents reaching Fermi's golden rule in the long-time limit. The novel diffusion approach essentially restores the correct asymptotics: As can be seen in the bottom row of Fig.~\ref{fig:G2_t50_G0}, aliased patterns are removed entirely. Instead, the imaginary part of $\mathcal{G}$, which is responsible for the dynamics of $G^\gtrless(t)$, only retains contributions near the resonance line, $\omega_{\mathbf{kpq}}^{\alpha\beta}=0$, as one expects from Fermi's golden rule.

\subsection{Dynamical $GW$ relaxation in 1D}\label{ssec:DynGW}
\begin{figure}[h]
    \centering
    \includegraphics[width=\linewidth]{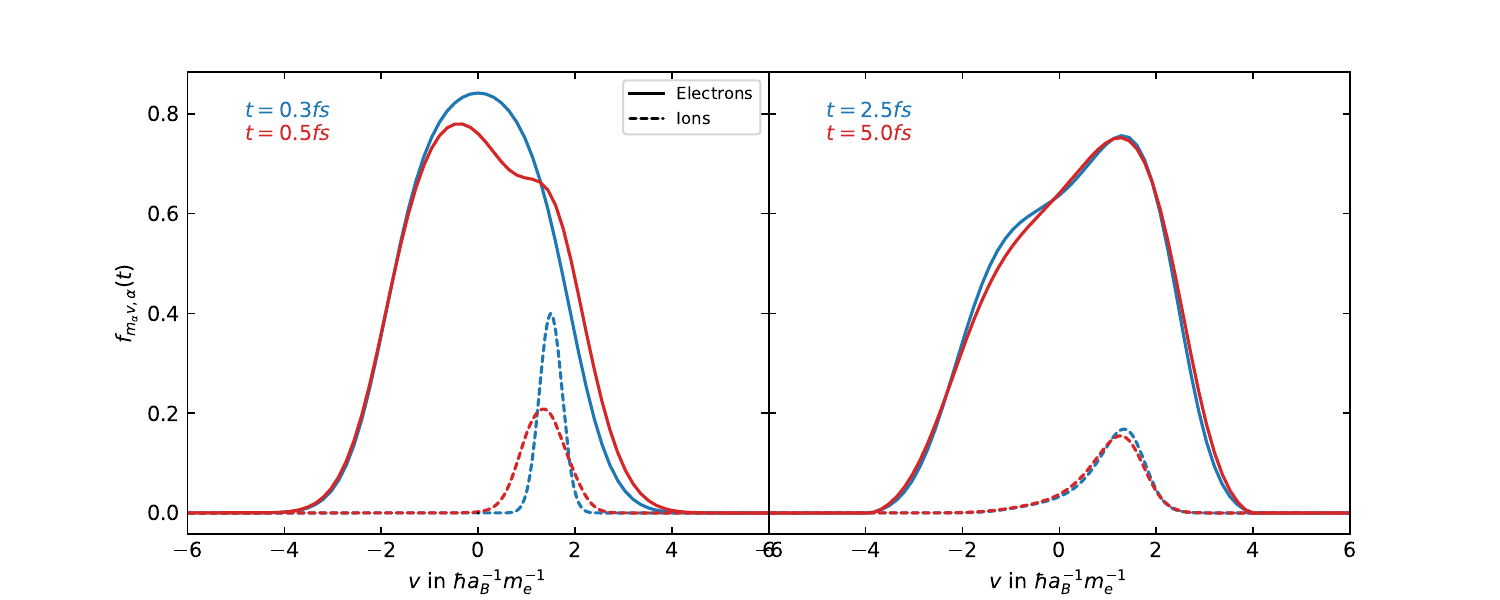}
    \caption{Test of the diffusion approach for a 1D ion stopping setup. At the initial time $t=0$, we initialize the electrons self-consistently on HF-level and subsequently switch on the interaction in the $\mathcal{G}$-equation (Adiabatic Switching). After that, at $t=0.3$fs, we introduce a Gaussian distribution of ions that represent a spatially uniform ion beam. Due to interaction, the two particle species exchange energy and momentum and slowly relax towards a common equilibrium state. 
    }
    \label{fig:TDDist}
\end{figure}
\begin{figure}[t]
    \centering
    \includegraphics[width=0.95\linewidth]{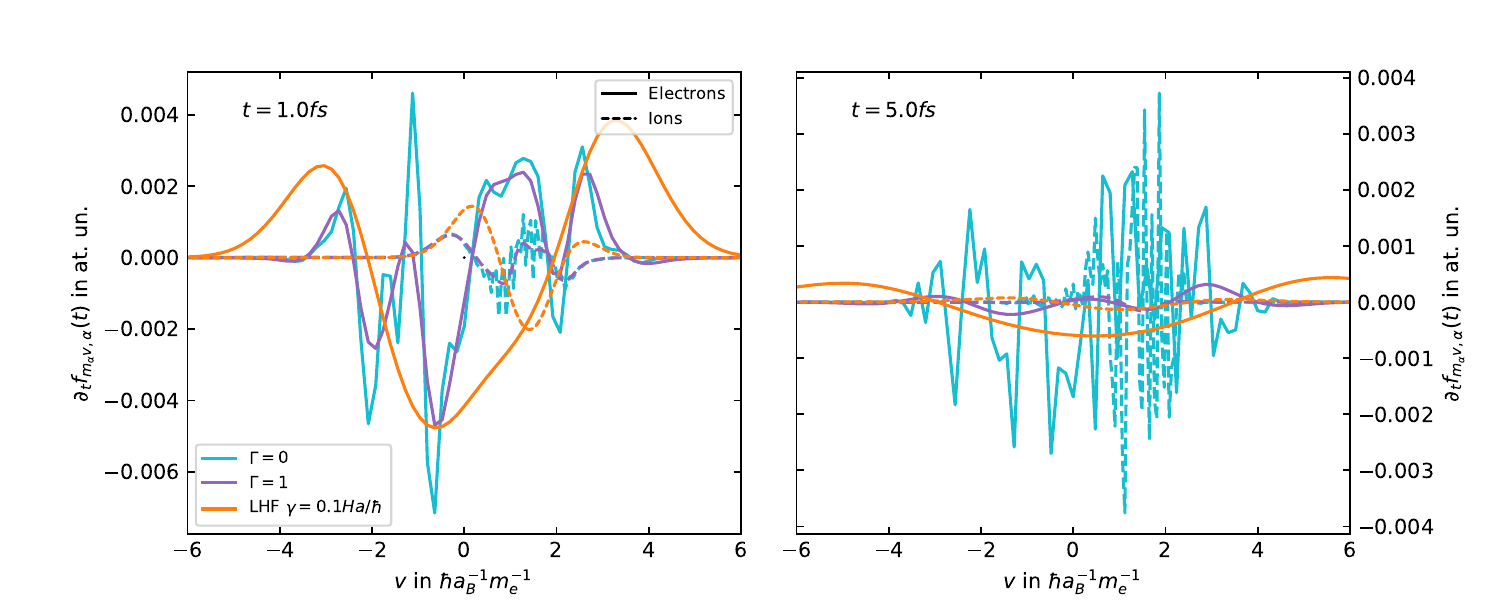}
    \caption{Time derivative of the distribution functions at the same times as in Fig. \ref{fig:TDDist}.  $\Gamma=0$ refers to the undamped HF-GKBA simulation and exhibits aliasing effects that grow with time. The Lorentzian HF-GKBA (LHF-GKBA) suppresses aliasing but violates energy conservation, cf. Fig.~\ref{fig:DynGWObs} and causes the occupation of states unphysically far above the Fermi edge. 
    }
    \label{fig:GWDynDeriv}
\end{figure}

Next we apply our diffusion approach to the $GW$ approximation and allow $G^\gtrless$ to evolve over time and no longer neglect exchange-energies in the two-particle commutator, i.e., we lift all simplifying restrictions that we previously made. 
Let us start with the original G1-G2 equations without diffusion term. 
The time-dependent distribution functions are shown in Fig. \ref{fig:TDDist}. At the initial time the electrons are uncorrelated and correlations are slowly introduced via the ``adiabatic switching'' procedure which produces an unperturbed correlated state at around $t=0.3 $fs. At this time the ion beam is introduced into the the system, and electrons and ions start colliding, exchanging momentum and energy. On average, the ions are stopped, but they also obtain more thermal energy, i.e., their distribution broadens. At the same time, the electron distribution changes its shape and develops a shoulder around $t=0.5 $fs which disappears in the course of the further evolution. At the end of the simulation both distributions are similar in shape, and the continuing relaxation towards Fermi distributions happens on a longer time scale. Fig. \ref{fig:GWDynDeriv} shows the time-derivative of the distribution functions. Aliasing manifests itself in the form of fluctuating spikes (as a function of momentum and time) of the time-derivative, if diffusion is turned off. 

\begin{figure}[t]
    \centering
    \includegraphics[width=0.95\linewidth]{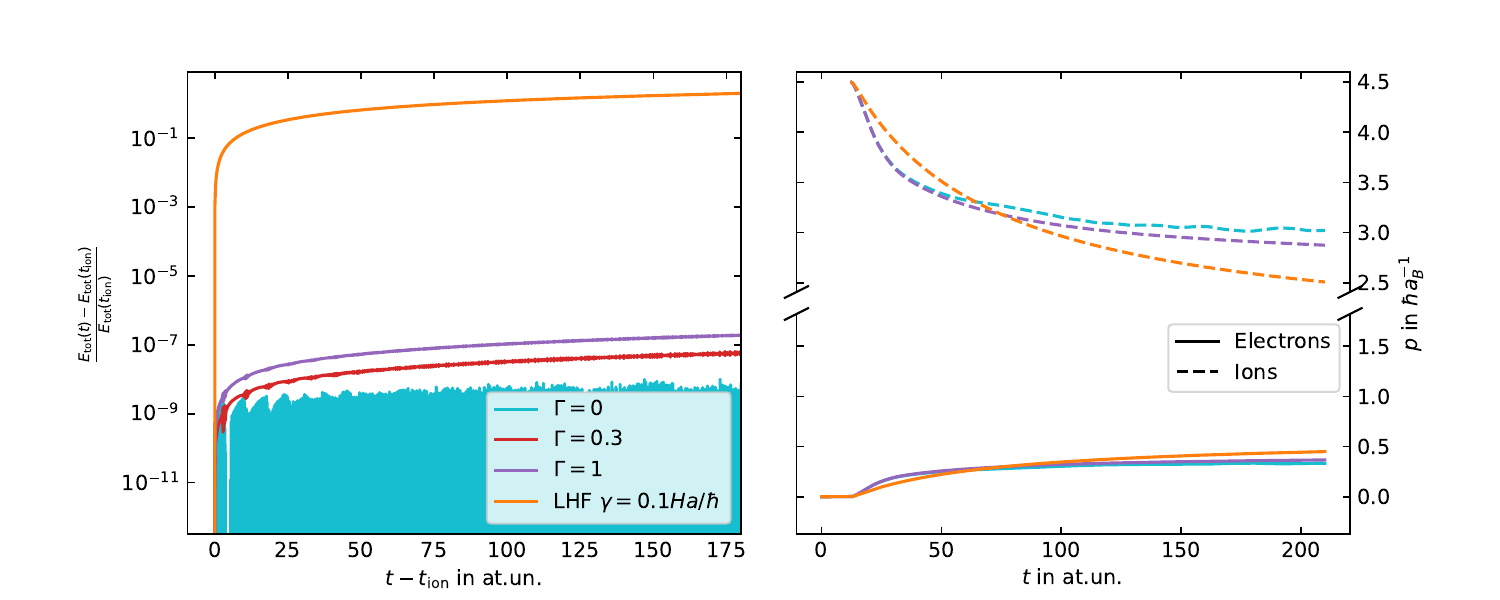}
    \caption{Left: total energy dynamics for three approximations corresponding to Fig.~\ref{fig:GWDynDeriv}. The HF-GKBA without diffusion ($\Gamma=0$) conserves energy. HF-GKBA with diffusion ($\Gamma=0.3$ and $\Gamma=1$) exhibits very good conservation properties as well. In contrast, the LHF-GKBA strongly violates the energy conservation. Right: Momentum transfer over time. The aliasing leads to a reduction and smaller oscillations in the momentum transfer. This is removed with $\Gamma=1$, and also by the LHF-GKBA. 
    }
    \label{fig:DynGWObs}
\end{figure}

Next, we present, for reference, results obtained with the LHF-GKBA, i.e. with an exponential damping constant $\gamma$, Eq.~\eqref{eq:hhf-damped}, where we choose $\gamma=0.1\,\mathrm{Ha}/\hbar$, as in Ref.~\onlinecite{makait_cpp_23}. This corresponds to the asymptotic damping in the high-density limit in SOA\cite{bonitz-etal.99epjb}. As Fig.~\ref{fig:GWDynDeriv} demonstrates, the LHF-GKBA (orange curves) successfully removes the aliasing. Finally, we repeat the simulations within the undamped HF-GKBA and the momentum diffusion turned on, using $\Gamma=1$. As can be seen from the purple curves in Fig.~\ref{fig:GWDynDeriv}, this approximation removes aliasing as effectively as does the LHF-GKBA. Both approximations are thus capable to perform long-time simulations. At the smae time, the quality of their results is very different as is obvious from the total energy conservation presented in Fig.~\ref{fig:DynGWObs}. While the LHF-approach significantly violates the energy conservation by more than 30\%, our novel  diffusion approach with $\Gamma=1$ exhibits a relative error that is below $10^{-6}$.
At the same time the momentum transfer (right figure part) is almost identical to the HF-GKBA without diffusion whereas the the
LHF-GKBA show significant deviations. These observations give a clear support of the diffusion approach within the HF-GKBA but rule out the LHF-GKBA.

\begin{figure}[h]
    \centering
    \includegraphics[width=0.9\linewidth]{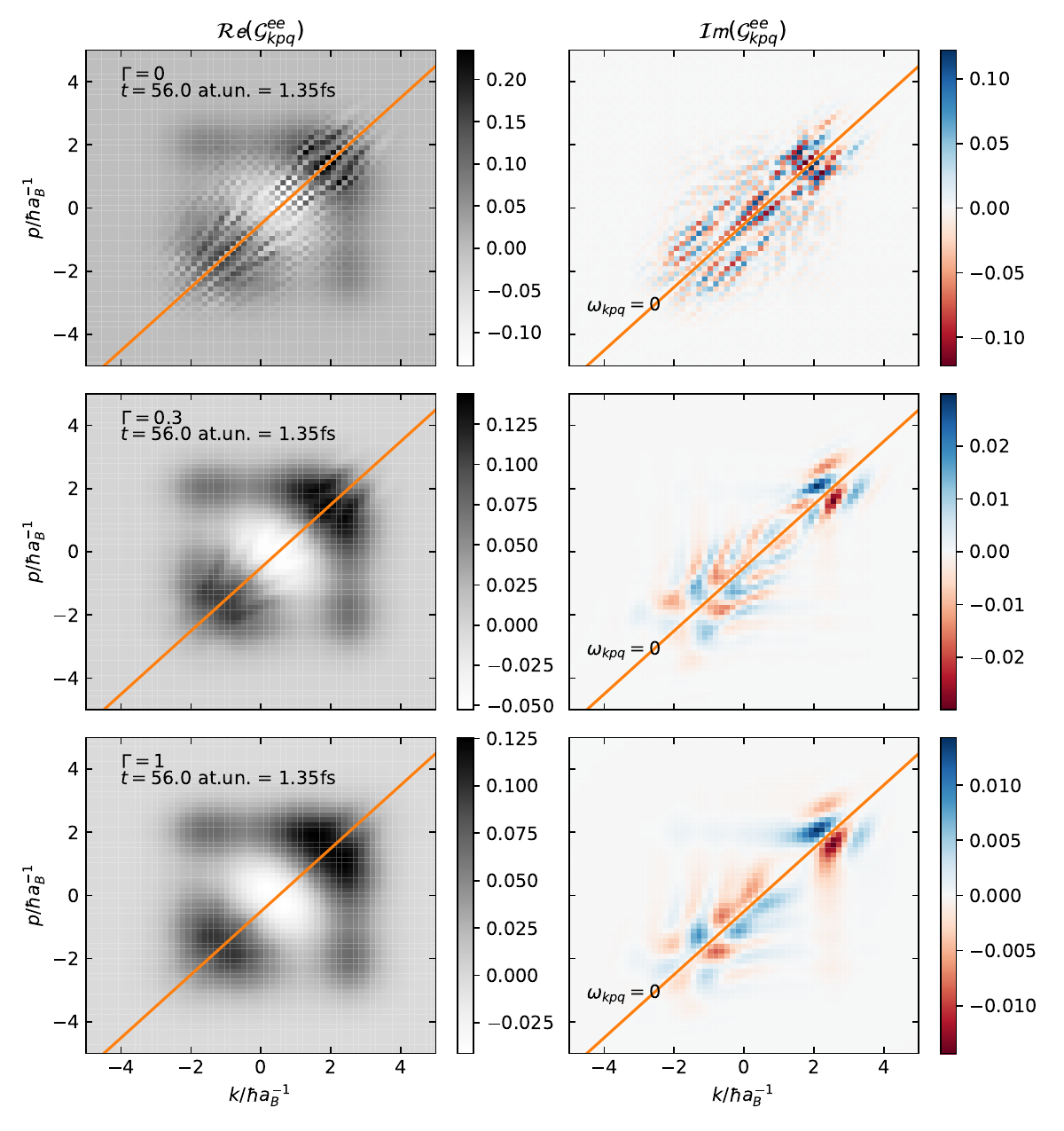}
    \caption{Electron-electron component of $\mathcal{G}$, for $q=0.5\,\hbar a_B^{-1}$ and three different rates of diffusion. The $\Gamma=0$ plot (upper row) clearly shows that at this time and the chosen $q$, $\mathcal{G}$ is at least near the aliasing time $\tau_\mathrm{alias}$. This issue is resolved by $\Gamma=0.3$ (middle row) and $\Gamma=1$ (bottom row). $\Gamma=1$ suppresses dense structures in $\mathcal{G}$ more than $\Gamma=0.3$, as one can directly see from the smaller number of peaks in the imaginary part. 
    }
    \label{fig:DynGW_G2_Gammas}
\end{figure}

\begin{figure}[h]
    \centering
    \includegraphics[width=0.9\linewidth]{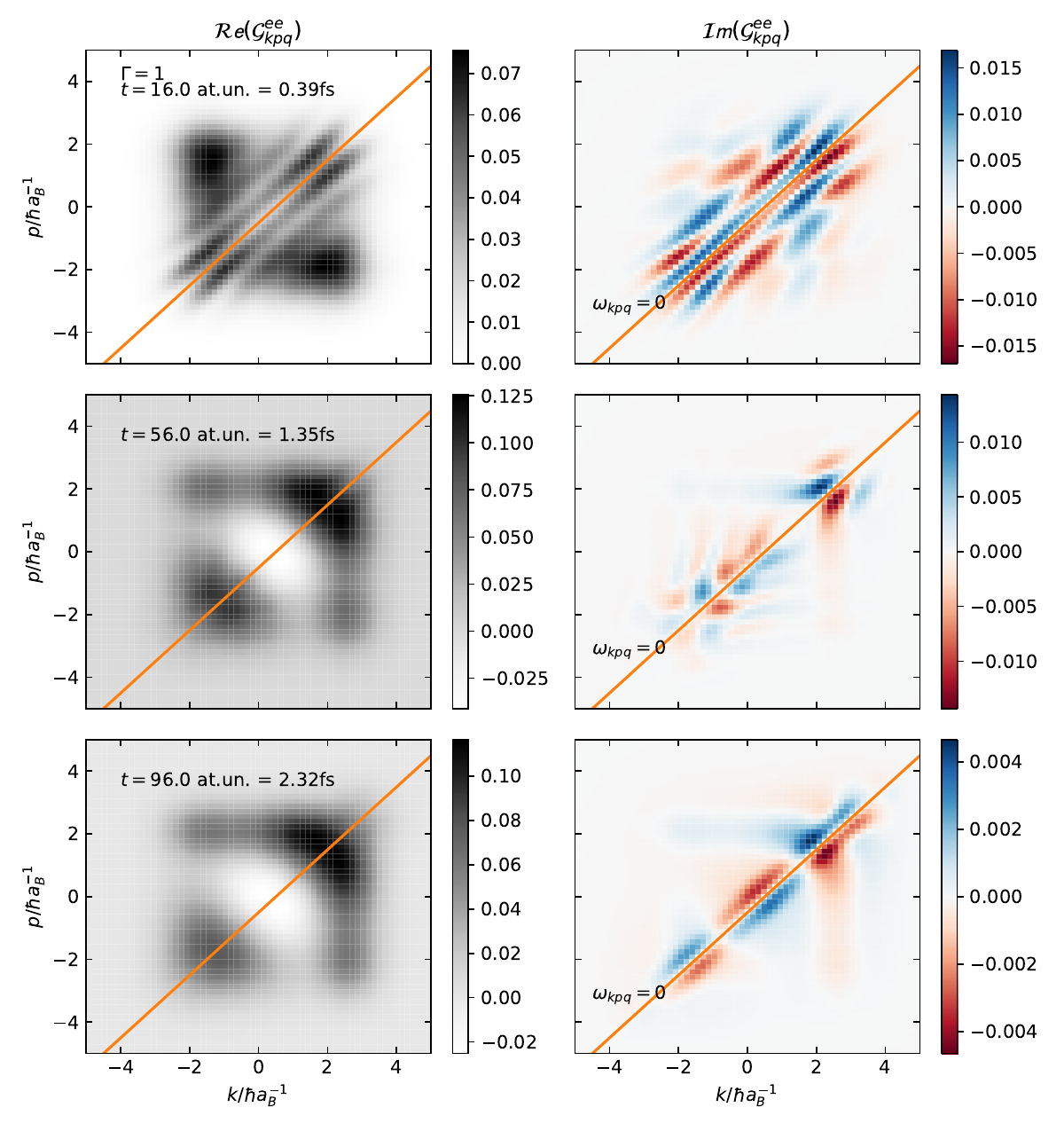}
    \caption{Electron-electron component of $\mathcal{G}$, for $q=0.5\,\hbar a_B^{-1}$ at $\Gamma=1$, for three different times. At earlier times (upper row) the density of oscillations builds up, but they can be well resolved on the grid. Compared to the simplified SOA case in Fig. \ref{fig:G2_t50_G0}, the pattern becomes more irregular in $GW$ with HF-GKBA. Aliasing never appears. The pattern converges into a quasi-stationary state at later times (bottom row) that changes only after larger changes of the single-particle quantities $G^\gtrless$. 
    }
    \label{fig:DynGW_G2_G1_ee}
\end{figure}

Let us now analyze in more detail the smoothing of the dynamics that is introduced by the diffusion terms and how it depends on $\Gamma$. A sensitive quantity is the two-particle electron correlation function, $\mathcal{G}^{ee}$, that is depicted ind Figs. \ref{fig:DynGW_G2_G1_ee} and \ref{fig:DynGW_G2_Gammas} for a few time points, as well as for three values  $\Gamma\in\{0,0.3,1\}$. We first study the influence of the parameter $\Gamma$ for a fixed time. The top panel of Figure~\ref{fig:DynGW_G2_Gammas} 
depicts the onset of aliasing, for the case without diffusion. Already a small diffusion amplitude, $\Gamma=0.3$ (middle row), completely removes aliasing. There are still weak oscillations in momentum space  but with a much larger wavelength compared to the case $\Gamma=0$. This trend is confirmed by choosing an approximatels three times larger parameter, $\Gamma=1$,  leading to a coarser structure of $\mathcal{G}$, cf. bottom row. Due to the excellent conservation properties that were demonstrated in Fig.~\ref{fig:DynGWObs}, this appears to be a reasonable choice.  
We, therefore, use the case $\Gamma=1$ in Fig.~\ref{fig:DynGW_G2_G1_ee} for an analysis of the time evolution of $\mathcal{G}$ within the diffusion approach. Interestingly, even though, at early times (top row)  $\mathcal{G}$ devlops weak precursors of aliasing, they are not giving rise to high frequency oscillations in momentum space and are reliably suppressed at later times. The same behavior is found for the electron-ion correlation function (not shown).
These results confirm the observations made for the simpler SOA case with the free GKBA, in  Sec.~\ref{ssec:QStatSOA}.

\section{Conclusions and outlook}\label{sec:4}
Aliasing is a fundamental problem in non-Markovian quantum kinetic simulations of discretized spatially uniform systems without sufficient damping. Even if the numerical scalings of the method itself (e.g. G1--G2 scheme) allows for arbitrarily long calculations, at some point in time the results will become unreliable. Prolonged calculations would require a finer $k$-point grid to postpone the emergence of aliasing to a later time. Previous approaches to removing the aliasing altogether involved introducing an artificial damping term (e.g. Lorentzian HF-GKBA). While they are in principle successful in removing the aliasing, these approaches severely violate conservation of total energy and give rise to a distorted spectral function. In this paper we introduced a fundamentally different approach. It introduces diffusion terms in mometum space and is reminiscent of a `coarse-graining' procedure. 
The diffusion apprach successfully suppresses aliasing and, at the same time, preserves total energy conservation. At early times -- when no aliasing is present and the $\mathcal
G$ is well represented on a grid -- the diffusive approach coincides with the original G1-G2 scheme.

The calculations, derivations and discussions have been carried out for a parabolic dispersion. An extension to systems where the interaction $w$ does not only depend on the transfer momentum $q$, such as in solids, is possible by adjusting the `weights' in the discrete Laplacian. Then, this approach is applicable to general uniform many-particle systems.
So far results were presented for quasi-one dimensional plasmas with SOA and GW selfenergies.
A preliminary analysis indicates that the diffusion approach might also allow for  efficient $GW$-HF-GKBA simulations of uniform systems in higher dimensions. For this case, until now, G1-G2 simulations are prevented by limitations on computer storage what does not allow to properly resolve $\mathcal{G}$. An alternative is to resort to a non-Makrovian formulation which will become efficient when the diffusion approach  is applied. This will be presented in a future publication. 

Due to the averaging and the involved information loss, the diffusion terms breaks the time reversal symmetry, even though no Markov limit has been performed, cf. Refs. \onlinecite{scharnke_jmp17,bonitz_cpp18}. How the long-time limit looks like and whether it differs from Fermi's golden rule, 
remains a question of ongoing research. Similarly, we expect that comparison with the thermodynamic asymptotics of the system will  allow to obtain indepent information on the optimal value of the free parameter $\Gamma$.

\section*{Acknowledgements}
We acknowledge fruitful discussions with E. Schroedter. This work has been supported by the Deutsche Forschungsgemeinschaft via grant BO1366/16.

\providecommand{\url}[1]{\texttt{#1}}
\providecommand{\urlprefix}{}
\providecommand{\foreignlanguage}[2]{#2}
\providecommand{\Capitalize}[1]{\uppercase{#1}}
\providecommand{\capitalize}[1]{\expandafter\Capitalize#1}
\providecommand{\bibliographycite}[1]{\cite{#1}}
\providecommand{\bbland}{and}
\providecommand{\bblchap}{chap.}
\providecommand{\bblchapter}{chapter}
\providecommand{\bbletal}{et~al.}
\providecommand{\bbleditors}{editors}
\providecommand{\bbleds}{eds: }
\providecommand{\bbleditor}{editor}
\providecommand{\bbled}{ed.}
\providecommand{\bbledition}{edition}
\providecommand{\bbledn}{ed.}
\providecommand{\bbleidp}{page}
\providecommand{\bbleidpp}{pages}
\providecommand{\bblerratum}{erratum}
\providecommand{\bblin}{in}
\providecommand{\bblmthesis}{Master's thesis}
\providecommand{\bblno}{no.}
\providecommand{\bblnumber}{number}
\providecommand{\bblof}{of}
\providecommand{\bblpage}{page}
\providecommand{\bblpages}{pages}
\providecommand{\bblp}{p}
\providecommand{\bblphdthesis}{Ph.D. thesis}
\providecommand{\bblpp}{pp}
\providecommand{\bbltechrep}{}
\providecommand{\bbltechreport}{Technical Report}
\providecommand{\bblvolume}{volume}
\providecommand{\bblvol}{Vol.}
\providecommand{\bbljan}{January}
\providecommand{\bblfeb}{February}
\providecommand{\bblmar}{March}
\providecommand{\bblapr}{April}
\providecommand{\bblmay}{May}
\providecommand{\bbljun}{June}
\providecommand{\bbljul}{July}
\providecommand{\bblaug}{August}
\providecommand{\bblsep}{September}
\providecommand{\bbloct}{October}
\providecommand{\bblnov}{November}
\providecommand{\bbldec}{December}
\providecommand{\bblfirst}{First}
\providecommand{\bblfirsto}{1st}
\providecommand{\bblsecond}{Second}
\providecommand{\bblsecondo}{2nd}
\providecommand{\bblthird}{Third}
\providecommand{\bblthirdo}{3rd}
\providecommand{\bblfourth}{Fourth}
\providecommand{\bblfourtho}{4th}
\providecommand{\bblfifth}{Fifth}
\providecommand{\bblfiftho}{5th}
\providecommand{\bblst}{st}
\providecommand{\bblnd}{nd}
\providecommand{\bblrd}{rd}
\providecommand{\bblth}{th}

\end{document}